%% file: sparse-beamforming.tex
\documentclass[draftclsnofoot, onecolumn, 12pt]{IEEEtran}

\usepackage{fixltx2e}
\usepackage{cite}
\usepackage{graphicx}
\usepackage{psfrag}
\usepackage{subfigure}
\usepackage{amsmath}
\usepackage{amssymb}
\usepackage{epsfig}
\usepackage{color}
\usepackage{epsf}
\usepackage{algorithmic}
\usepackage{algorithm}
\usepackage{epsfig}
\usepackage{stmaryrd}
\usepackage{textcomp}
\usepackage{setspace}
\usepackage[update, prepend]{epstopdf}
\input{input.tex}
\DeclareMathOperator*{\argmax}{arg\,max}
\DeclareMathOperator*{\argmin}{arg\,min}
\def\Nt{{N_\mathrm{t}}}         %
\def\Nr{{N_\mathrm{r}}}         %
\def\Ns{{N_\mathrm{s}}}         %
\def\NtRF{{N_\mathrm{t}^\mathrm{RF}}}         %
\def\NrRF{{N_\mathrm{r}^\mathrm{RF}}}

\def\Fopt{\bF_\mathrm{opt}} 
 
\def\Fbb{\bF_\mathrm{BB}} 
\def\Frf{\bF_\mathrm{RF}} 
\def\Wbb{\bW_\mathrm{BB}} 
\def\Wrf{\bW_\mathrm{RF}} 
\def\Wmmse{\bW_\mathrm{MMSE}} 
\def\SigmaH{\boldsymbol{\Sigma}} 

\def\sigman{\sigma_\mathrm{n}}  
\def\Ncl{N_\mathrm{cl}}  
\def\Nray{N_\mathrm{ray}}
\def\LambdaR{\Lambda_\mathrm{r}}
\def\LambdaT{\Lambda_\mathrm{t}}
\def\ar{\ba_\mathrm{r}}
\def\at{\ba_\mathrm{t}}
\def\aULA{\ba_\mathrm{ULAy}}
\def\aUPA{\ba_\mathrm{UPA}}
\def\Phit{\phi^\mathrm{t}}
\def\Phir{\phi^\mathrm{r}}
\def\Thetat{\theta^\mathrm{t}}
\def\Thetar{\theta^\mathrm{r}}

\def\Nphi{{N_\phi}}            %
\def\Ntheta{{N_\theta}}        %

\IEEEoverridecommandlockouts
\begin{document}
%
\title{Spatially Sparse Precoding in Millimeter Wave MIMO Systems~\thanks{This work was done while the first author was with Samsung Research America - Dallas. The authors at The University of Texas at Austin were supported in part by the Army Research Laboratory contract W911NF-10-1-0420 and National Science Foundation grant 1218338. This work has appeared in part at the 2012 IEEE International Communications Conference (ICC).}}

\author{\IEEEauthorblockN{Omar~El~Ayach, Sridhar Rajagopal, Shadi Abu-Surra, Zhouyue Pi, and Robert~W.~Heath,~Jr.\footnote{Omar El Ayach and Robert Heath are with The University of Texas at Austin, Austin, TX 78712 USA (Email: {oelayach, rheath}@utexas.edu). Sridhar Rajagopal, Shadi Abu-Surra, and Zhouyue Pi are with Samsung Research America - Dallas, Richardson, TX, 75082 USA (Email: \{sasurra, srajagop, zpi\}@sta.samsung.com)}}}


%


\maketitle
\vspace{-0pt}

\begin{abstract}

Millimeter wave (mmWave) signals experience orders-of-magnitude more pathloss than the microwave signals currently used in most wireless applications. MmWave systems must therefore leverage large antenna arrays, made possible by the decrease in wavelength, to combat pathloss with beamforming gain. Beamforming with multiple data streams, known as precoding, can be used to further improve mmWave spectral efficiency. Both beamforming and precoding are done digitally at baseband in traditional multi-antenna systems. The high cost and power consumption of mixed-signal devices in mmWave systems, however, make analog processing in the RF domain more attractive. This hardware limitation restricts the feasible set of precoders and combiners that can be applied by practical mmWave transceivers. In this paper, we consider transmit precoding and receiver combining in mmWave systems with large antenna arrays. We exploit the spatial structure of mmWave channels to formulate the precoding/combining problem as a sparse reconstruction problem. Using the principle of basis pursuit, we develop algorithms that accurately approximate optimal unconstrained precoders and combiners such that they can be implemented in low-cost RF hardware. 
We present numerical results on the performance of the proposed algorithms and show that they allow mmWave systems to approach their unconstrained performance limits, even when transceiver hardware constraints are considered.
\end{abstract}

\section{Introduction}

The capacity of wireless networks has thus far scaled with the increasing data traffic, primarily due to improved area spectral efficiency (bits/s/Hz/m\textsuperscript{2})~\cite{ghosh2010lte}. A number of physical layer enhancements such as multiple antennas, channel coding, and interference coordination, as well as the general trend toward network densification have all been instrumental in achieving this efficiency~\cite{ghosh2010lte, lopez2011enhanced}. Since there seems to be little scope for further gains at the physical layer, and since the widespread deployment of heterogeneous networks is not without challenges~\cite{damnjanovic2011survey}, these techniques alone may not be sufficient to meet future traffic demands. As a result, increasing the spectrum available for commercial wireless systems, potentially by exploring new less-congested spectrum bands, is a promising solution to increase network capacity.

Millimeter wave (mmWave) communication, for example, has enabled gigabit-per-second data rates in indoor wireless systems~\cite{yong2007overview, daniels200760} and fixed outdoor systems~\cite{papazian1997study}. More recently, advances in mmWave hardware \cite{doan2004design} and the potential availability of spectrum has encouraged the wireless industry to consider mmWave for outdoor cellular systems~\cite{pi2011introduction, hendrantoro2002use}. A main differentiating factor in mmWave communication is that the ten-fold increase in carrier frequency, compared to the current majority of wireless systems, implies that mmWave signals experience an orders-of-magnitude increase in free-space pathloss. An interesting redeeming feature in mmWave systems, however, is that the decrease in wavelength enables packing large antenna arrays at both the transmitter and receiver. Large arrays can provide the beamforming gain needed to overcome pathloss and establish links with reasonable signal-to-noise ratio (SNR). Further, large arrays may enable precoding multiple data streams which could improve spectral efficiency and allow systems to approach capacity~\cite{torkildson2009millimeter, madhow-four-channel-spatial-multiplexing}.

While the fundamentals of precoding are the same regardless of carrier frequency, signal processing in mmWave systems is subject to a set of non-trivial practical constraints. For example, traditional multiple-input multiple-output (MIMO) processing is often performed digitally at baseband, which enables controlling both the signal's phase and amplitude. Digital processing, however, requires dedicated baseband and RF hardware for each antenna element. Unfortunately, the high cost and power consumption of mmWave mixed-signal hardware precludes such a transceiver architecture at present, and forces mmWave systems to rely heavily on analog or RF processing \cite{doan2004design, pi2011introduction}. Analog precoding is often implemented using phase shifters~\cite{pi2011introduction, doan2004design, valdes2010fully} which places constant modulus constraints on the elements of the RF precoder. Several approaches have been considered for precoding in such low-complexity transceivers~\cite{sanayei2004antenna, molisch2005capacity, gorokhov2003receive, xu2009analysis, wang2009beam, sayeed2010continuous, sayeed2013mmWave, molisch2004fft, love-heath-EGT, molisch-channel-statistics-preprocessing, zheng2007mimo, nsenga2010mixed, gholam2011beamforming, pi2012optimal, molisch-variable-phase-shift, venkateswaran2010analog}. The work in \cite{sanayei2004antenna, molisch2005capacity, gorokhov2003receive} considers antenna (or antenna subset) selection which has the advantage of replacing phase shifters with even simpler analog switches. Selection, however, provides limited array gain and performs poorly in correlated channels such as those experienced in mmWave~\cite{xu2009analysis}. To improve performance over correlated channels, the work \cite{wang2009beam, sayeed2010continuous, sayeed2013mmWave, molisch2004fft} considers beam steering solutions in which phase shifters or discrete lens arrays are used to optimally orient an array's response in space, potentially based on statistical channel knowledge. The strategies in \cite{wang2009beam, sayeed2010continuous, sayeed2013mmWave, molisch2004fft}, however, are in general suboptimal since beam steering alone cannot perfectly capture the channels dominant eigenmodes. The work in \cite{love-heath-EGT, molisch-channel-statistics-preprocessing, zheng2007mimo, nsenga2010mixed, gholam2011beamforming, pi2012optimal} develops iterative precoding algorithms for systems that leverage analog processing, and \cite{molisch-variable-phase-shift} further proposes simple analytical solutions. Further hardware limitations have also been considered in \cite{venkateswaran2010analog}, for example, which focuses on analog receiver processing with only quantized phase control and finite-precision analog-to-digital converters. The work in \cite{love-heath-EGT, molisch-channel-statistics-preprocessing, zheng2007mimo, nsenga2010mixed, gholam2011beamforming, pi2012optimal, molisch-variable-phase-shift, venkateswaran2010analog}, however, is not specialized to mmWave MIMO systems with large antenna arrays. Namely, the work in \cite{love-heath-EGT, molisch-channel-statistics-preprocessing, zheng2007mimo, nsenga2010mixed, gholam2011beamforming, pi2012optimal, molisch-variable-phase-shift, venkateswaran2010analog} does not leverage the structure present in mmWave MIMO channels and adopts models that do not fully capture the effect of limited mmWave scattering and large tightly-packed arrays~\cite{smulders1997characterisation, xu2002spatial, spencer-channel-model}.

In this paper, we focus on the precoding insight and solutions that can be derived from jointly considering the following three factors: (i) precoding with RF hardware constraints, (ii) the use of large antenna arrays, and (iii) the limited scattering nature of mmWave channels. We consider single-user precoding for a practical transceiver architecture in which a large antenna array is driven by a limited number of transmit/receive chains~\cite{pi2011introduction,torkildson2009millimeter, madhow-four-channel-spatial-multiplexing,samsung-practical-sdma-60GHz}. In such a system, transmitters have the ability to apply high-dimensional (tall) RF precoders, implemented via analog phase shifters, followed by low-dimensional (small) digital precoders that can be implemented at baseband. We adopt a realistic clustered channel model that captures both the limited scattering at high frequency and the antenna correlation present in large tightly-packed arrays~\cite{smulders1997characterisation, spencer-channel-model, xu2002spatial}. 

We exploit the sparse-scattering structure of mmWave channels to formulate the design of hybrid RF/baseband precoders as a sparsity constrained matrix reconstruction problem~\cite{markovsky2007overview, tropp-signal-recovery, tropp2004greed, rebollo2002optimized, tibshirani1996regression, boyd2004convex}. Initial results on this precoding approach were presented in \cite{ayach2012low}. In this paper, we formalize the mmWave precoding problem and show that, instead of directly maximizing mutual information, near-optimal hybrid precoders can be found via an optimization that resembles the problem of sparse signal recovery with multiple measurement vectors, also known as the simultaneously sparse approximation problem~\cite{cotter2005sparse, tropp2006algorithmspart1, tropp2006algorithmspart2, chen2006theoretical}. We thus provide an algorithmic precoding solution based on the concept of orthogonal matching pursuit~\cite{rebollo2002optimized, tropp-signal-recovery, wright2009sparse}. The algorithm takes an optimal unconstrained precoder as input and approximates it as linear combination of beam steering vectors that can be applied at RF (and combined digitally at baseband). Further, we extend this sparse precoding approach to receiver-side processing and show that designing hybrid minimum mean-square error (MMSE) combiners can again be cast as a simultaneously sparse approximation problem and solved via basis pursuit~\cite{michaeli2007constrained, michaeli2008constrained}. We argue that, in addition to providing practical near-optimal precoders, the proposed framework is particularly amenable for limited feedback operation and is thus not limited to genie-aided systems with perfect transmitter channel knowledge~\cite{love-heath-limited-feedback}. The generated precoders can be efficiently compressed using simple scalar quantizers (for the arguments of the beam steering vectors) and low-dimensional Grassmannian subspace quantizers (used to quantize the baseband precoder)~\cite{love-heath-limited-feedback, love-heath-grassmannian-beamforming, roh2006design}. We briefly describe the construction of the limited feedback codebooks required, but defer the analysis of limited feedback performance to future work. Finally, we present simulation results on the performance of the proposed strategy and show that it allows mmWave systems to approach their unconstrained performance limits even when practical transceiver constraints are considered.


We use the following notation throughout this paper: $\bA$ is a matrix; $\ba$ is a vector; $a$ is a scalar; $\bA^{(i)}$ is the $i^{th}$ column of $\bA$; $(\cdot)^T$ and $(\cdot)^*$ denote transpose and conjugate transpose respectively; $\|\bA\|_{F}$ is the Frobenius norm of $\bA$, $\mathrm{tr}(\bA)$ is its trace and $|\bA|$ is its determinant; $\|\ba\|_p$ is the $p$-norm of $\ba$; $[\bA\ |\ \bB]$ denotes horizontal concatenation; $\mathrm{diag}(\bA)$ is a vector formed by the diagonal elements of $\bA$; $\bI_N$ is the $N\times N$ identity matrix; $\mathbf{0}_{M\times N}$ is the $M \times N$ all-zeros matrix; $\mathcal{CN}(\ba;\bA)$ is a complex Gaussian vector with mean $\ba$ and covariance matrix $\bA$. Expectation is denoted by $\bbE[\cdot]$ and the real part of a variable is denoted by $\Re\left\{\cdot \right\}$.

\section{System Model} \label{sec:SysModelMain}

In this section, we present the mmWave signal and channel model considered in this paper.

\subsection{System Model} \label{sec:SystemModel}

\begin{figure} [t]
  \centering
  \includegraphics[width=5.5in]{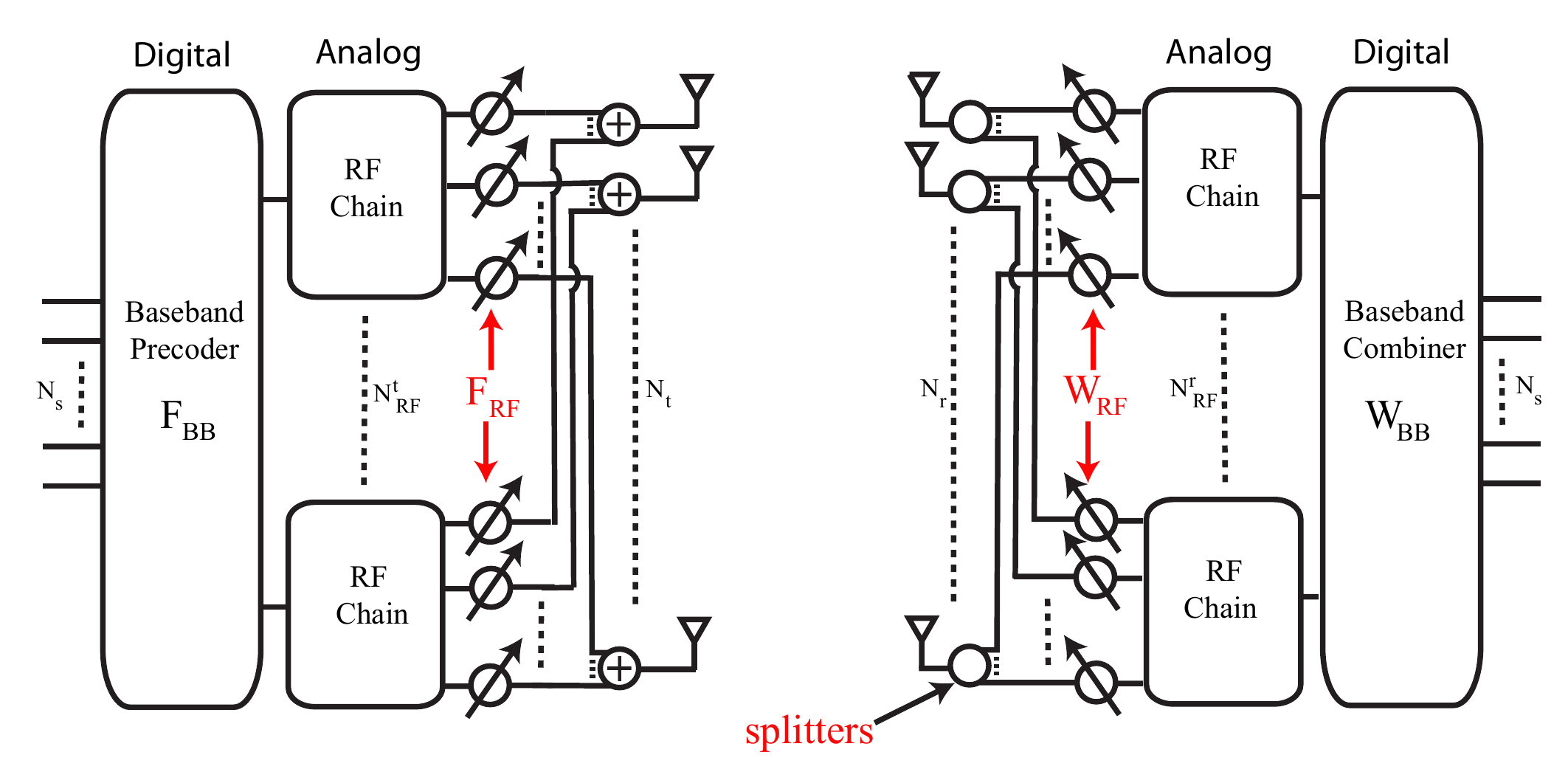}
  \caption{Simplified hardware block diagram of mmWave single user system with digital baseband precoding followed by constrained radio frequency precoding implemented using RF phase shifters.}
  \label{fig:MmWaveSystemModel}
  \vspace{-0.2in}
\end{figure}

Consider the single-user mmWave system shown in Fig. \ref{fig:MmWaveSystemModel} in which a transmitter with $\Nt$ antennas communicates $\Ns$ data streams to a receiver with $\Nr$ antennas~\cite{samsung-practical-sdma-60GHz}. To enable multi-stream communication, the transmitter is equipped with $\NtRF$ transmit chains such that $\Ns \leq \NtRF \leq \Nt$. This hardware architecture enables the transmitter to apply an $\NtRF \times \Ns$ baseband precoder $\Fbb$ using its $\NtRF$ transmit chains, followed by an $\Nt \times \NtRF$ RF precoder $\Frf$ using analog circuitry. The discrete-time transmitted signal is therefore given by $\bx=\Frf\Fbb\bs$
where $\bs$ is the $\Ns \times 1$ symbol vector such that $\bbE\left[\bs\bs^*\right]=\frac{1}{\Ns}\bI_{\Ns}$. Since $\Frf$ is implemented using analog phase shifters, its elements are constrained to satisfy $(\bF^{(i)}_\mathrm{RF}\bF^{(i)*}_\mathrm{RF})_{\ell,\ell}=\Nt^{-1}$, where $(\cdot)_{\ell,\ell}$ denotes the $\ell^{th}$ diagonal element of a matrix, i.e., all elements of $\Frf$ have equal norm. The transmitter's total power constraint is enforced by normalizing $\Fbb$ such that $\|\Frf\Fbb\|_{F}^2=\Ns$; no other hardware-related constraints are placed on the baseband precoder. 

For simplicity, we consider a narrowband block-fading propagation channel as in \cite{torkildson2009millimeter, sayeed2013mmWave, pi2012optimal, samsung-practical-sdma-60GHz}, which yields a received signal
\begin{equation}
\by=\sqrt{\rho}\bH\Frf\Fbb\bs+\bn,
\label{eqn:RxSignal}
\end{equation}
where $\by$ is the $\Nr \times 1$ received vector, $\bH$ is the $\Nr \times \Nt$ channel matrix such that $\bbE\left[\|\bH\|_{F}^2\right]=\Nt\Nr$, $\rho$ represents the average received power, and $\bn$ is the vector of i.i.d $\mathcal{CN}(0,\sigman^2)$ noise. In writing (\ref{eqn:RxSignal}), we implicitly assume perfect timing and frequency recovery. Moreover, to enable precoding, we assume that the channel $\bH$ is known perfectly and instantaneously to both the transmitter and receiver. In practical systems, channel state information (CSI) at the receiver can be obtained via training~\cite{wang2009beam, alkhateebhybrid} and subsequently shared with the transmitter via limited feedback~\cite{love-heath-limited-feedback}; an efficient limited feedback strategy is presented in Section \ref{sec:LimitedFeedback}. Techniques for efficient mmWave channel estimation, and a rigorous treatment of frequency selective mmWave channels, are still an ongoing topic of research.

The receiver uses its $\Ns \leq \NrRF \leq \Nr$ RF chains and its analog phase shifters to obtain the post-processing received signal
\begin{equation}
\widetilde{\by}=\sqrt{\rho}\Wbb^*\Wrf^* \bH\Frf\Fbb\bs+\Wbb^*\Wrf^*\bn,
\label{eqn:CombinedRxSignal}
\end{equation}
where $\Wrf$ is the $\Nr \times \NrRF$ RF combining matrix and $\Wbb$ is the $\NrRF \times \Ns$ baseband combining matrix. Similarly to the RF precoder, $\Wrf$ is implemented using phase shifters and therefore is such that $(\bW^{(i)}_\mathrm{RF}\bW^{(i)*}_\mathrm{RF})_{\ell,\ell}=\Nr^{-1}$. When Gaussian symbols are transmitted over the mmWave channel, the spectral efficiency achieved is given by~\cite{goldsmith2003capacity}
\begin{align}
\begin{split}
R=\log_2\left(\left| \bI_{\Ns} + \frac{\rho}{\Ns}\bR_\mathrm{n}^{-1}  \Wbb^*\Wrf^*\bH\Frf\Fbb
 \Fbb^*\Frf^*\bH^*\Wrf\Wbb \right|\right),
\label{eqn:SpectralEfficiency}
\end{split}
\end{align}
where $\bR_\mathrm{n}=\sigman^2 \Wbb^*\Wrf^*\Wrf\Wbb$ is the noise covariance matrix after combining.

\subsection{Channel Model} \label{sec:ChannelModel}

The high free-space pathloss that is a characteristic of mmWave propagation leads to limited spatial selectivity or scattering. Similarly, the large tightly-packed antenna arrays that are characteristic of mmWave transceivers lead to high levels of antenna correlation. This combination of tightly packed arrays in sparse scattering environments makes many of the statistical fading distributions used in traditional MIMO analysis inaccurate for mmWave channel modeling. For this reason, we adopt a narrowband clustered channel representation, based on the extended Saleh-Valenzuela model, which allows us to accurately capture the mathematical structure present in mmWave channels~\cite{xu2002spatial, smulders1997characterisation, ieee802153c, raghavan2006multi}.

Using the clustered channel model, the matrix channel $\bH$ is assumed to be a sum of the contributions of $\Ncl$ scattering clusters, each of which contribute $\Nray$ propagation paths to the channel matrix $\bH$. Therefore, the discrete-time narrowband channel $\bH$ can be written as 
\begin{equation}
\bH=\sqrt{\frac{\Nt\Nr}{\Ncl\Nray}}\sum\limits_{i=1}^{\Ncl}\sum\limits_{\ell=1}^{\Nray}\alpha_{i\ell}\LambdaR(\Phir_{i\ell}, \Thetar_{i\ell})\LambdaT(\Phit_{i\ell},\Thetat_{i\ell}) \ar(\Phir_{i\ell},\ \Thetar_{i\ell})\at(\Phit_{i\ell},\ \Thetat_{i\ell})^*,
\label{eqn:ClusteredChannel}
\end{equation}
where $\alpha_{i\ell}$ is the complex gain of the $\ell^\text{th}$ ray in the $i^\text{th}$ scattering cluster, whereas $\Phir_{i\ell}$ ($\Thetar_{i\ell}$) and $\Phit_{i\ell}$ ($\Thetat_{i\ell}$) are its azimuth (elevation) angles of arrival and departure respectively. The functions $\LambdaT(\Phit_{i\ell},\Thetat_{i\ell})$ and $\LambdaR(\Phir_{i\ell}, \Thetar_{i\ell})$ represent the transmit and receive \emph{antenna element gain} at the corresponding angles of departure and arrival. Finally, the vectors $\ar(\Phir_{\ell},\Thetar_{\ell})$ and $\at(\Phit_{i\ell},\Thetat_{i\ell})$ represent the normalized receive and transmit array response vectors at an azimuth (elevation) angle of $\Phir_{i\ell}$ ($\Thetar_{i\ell}$) and $\Phit_{i\ell}$ ($\Thetat_{i\ell}$) respectively.

In Section \ref{sec:Simulations}, we assume that $\alpha_{i\ell}$ are i.i.d.  $\mathcal{CN} (0, \sigma_{\alpha,i}^2)$ where $\sigma_{\alpha,i}^2$ represents the average power of the $i^\text{th}$ cluster. The average cluster powers are such that $\textstyle{\sum_{i=1}^{\Ncl} \sigma_{\alpha,i}^2=\gamma}$ where $\gamma$ is a normalization constant that satisfies $\bbE\left[\|\bH\|_{F}^2\right]=\Nt\Nr$~\cite{xu2002spatial}.
The $\Nray$ azimuth and elevation angles of departure, $\Phit_{i\ell}$ and $\Thetat_{i\ell}$, within the cluster $i$ are assumed to be randomly distributed with a uniformly-random mean cluster angle of $\Phit_i$ and $\Thetat_i$ respectively, and a constant angular spread (standard deviation) of $\sigma_{\Phit}$ and $\sigma_{\Thetat}$ respectively. The azimuth and elevation angles of arrival, $\Phir_{i\ell}$ and $\Thetar_{i\ell}$, are again randomly distributed with mean cluster angles of $(\Phir_i,\ \Thetar_i)$ and angular spreads $(\sigma_{\Phit},\ \sigma_{\Thetar})$. While a variety of distributions have been proposed for the angles of arrival and departure in clustered channel models, the Laplacian distribution has been found to be a good fit for a variety of propagation scenarios~\cite{forenza2007simplified}, and will thus be adopted in the numerical results of Section~\ref{sec:Simulations}. 
Similarly, a number of parametrized mathematical models have been proposed for the functions $\LambdaT(\Phit_{i\ell},\Thetat_{i\ell})$ and $\LambdaR(\Phir_{i\ell}, \Thetar_{i\ell})$. For example, if the transmitter's antenna elements are modeled as being ideal sectored elements~\cite{singh2011interference}, $\LambdaT(\Phit_{i\ell},\Thetat_{i\ell})$ would be given by
\begin{equation}
\LambdaT(\Phit_{i\ell},\Thetat_{i\ell})= \left\{ \begin{array}{cc} 1 &  \quad \forall \Phit_{i\ell}\in [\Phit_\mathrm{min}, \Phit_\mathrm{max}],\ \forall \Thetat_{i\ell}\in [\Thetat_\mathrm{min}, \Thetat_\mathrm{max}], \\ 0 & \mathrm{otherwise}, \end{array} \right.
\label{eqn:AntennaElementGain}
\end{equation}
where we have assumed unit gain over the sector defined by $\Phit_{\ell}\in [\Phit_\mathrm{min}, \Phit_\mathrm{max}]$ and $\Thetat_{\ell}\in [\Thetat_\mathrm{min}, \Thetat_\mathrm{max}]$ without loss of generality.
The receive antenna element gain $\LambdaR(\Phir_{i\ell}, \Thetar_{i\ell})$ is defined similarly over the azimuth sector $\Phir_{i\ell}\in [\Phir_\mathrm{min}, \Phir_\mathrm{max}]$ and elevation sector $\Thetar_{i\ell}\in [\Thetar_\mathrm{min}, \Thetar_\mathrm{max}]$. 

The array response vectors $\at(\Phit_{i\ell},\Thetat_{i\ell})$ and $\ar(\Phir_{\ell},\Thetar_{\ell})$ are a function of the transmit and receiver antenna array structure only, and are thus independent of the antenna element properties. While the algorithms and results derived in the remainder of this paper can be applied to arbitrary antenna arrays, we give the following two illustrative examples of commonly-used antenna arrays for completeness. For an $N$-element uniform linear array (ULA) on the $y$-axis, the array response vector can be written as~\cite{balanis1997antenna}
\begin{equation}
\aULA(\phi)=\frac{1}{\sqrt{N}}\left[1,\ e^{jkd\sin(\phi)},\ e^{j2kd\sin(\phi)},\ \hdots,\  e^{j(N-1)kd\sin(\phi)}\right]^T,
\label{eqn:ULA}
\end{equation}
where $k=\frac{2\pi}{\lambda}$ and $d$ is the inter-element spacing. Note that we do not include $\theta$ in the arguments of $\aULA$ as the array's response is invariant in the elevation domain. In the case of a uniform planar array (UPA) in the $yz$-plane with $W$ and $H$ elements on the $y$ and $z$ axes respectively, the array response vector is given by \cite{balanis1997antenna}
\begin{align}
\begin{split}
\aUPA(\phi, \theta)= \frac{1}{\sqrt{N}} \left(\right. \left[\right. & 1,\ \hdots,\  e^{jkd(m\sin(\phi)\sin(\theta)+n\cos(\theta))},\ \hdots, \\
& \hspace{130pt} \hdots,\ e^{jkd((W-1)\sin(\phi)sin(\theta)+(H-1)\cos(\theta))}\left. \right]\left.\right)^T,
\label{eqn:UPA}
\end{split}
\end{align}
where $0\leq m <W$ and $0\leq n <H$ are the $y$ and $z$ indices of an antenna element respectively and the antenna array size is $N=WH$. Considering uniform planar arrays is of interest in mmWave beamforming since they (i) yield smaller antenna array dimensions, (ii) facilitate packing more antenna elements in a reasonably-sized array, and (iii) enable beamforming in the elevation domain (also known as 3D beamforming).

\section{Spatially Sparse Precoding for the Single User mmWave Channel} \label{sec:SparsePrecoding}

We seek to design hybrid mmWave precoders $(\Frf,\Fbb)$ that maximize the spectral efficiency expression in (\ref{eqn:SpectralEfficiency}). Directly maximizing (\ref{eqn:SpectralEfficiency}), however, requires a joint optimization over the four matrix variables $(\Frf,\Fbb,\Wrf,\Wbb)$. Unfortunately, finding global optima for similar constrained joint optimization problems is often found to be intractable~\cite{palomar2003joint, palomar2006tutorial}. In the case of mmWave precoding, the non-convex constraints on $\Frf$ and $\Wrf$ makes finding an exact solution unlikely. To simplify transceiver design, we temporarily decouple the joint transmitter-receiver optimization problem and focus on the design of the hybrid precoders $\Frf\Fbb$. Therefore, in lieu of maximizing spectral efficiency, we design $\Frf\Fbb$ to maximize the mutual information achieved by Gaussian signaling over the mmWave channel
\begin{equation}
\mathcal{I}(\Frf, \Fbb)=\log_2\left(\left| \bI_{\Ns} + \frac{\rho}{\Ns\sigman^2} \bH \Frf\Fbb \Fbb^*\Frf^*\bH^* \right|\right).
\label{eqn:MutualInformation}
\end{equation} 
We note here that abstracting receiver operation, and focusing on mutual information instead of the spectral efficiency expression in (\ref{eqn:SpectralEfficiency}), effectively amounts to assuming that the receiver can perform optimal nearest-neighbor decoding based on the $\Nr$-dimensional received signal $\by$. Unfortunately, such a decoder is impossible to realize with practical mmWave systems in which decoders do not have access to the $\Nr$-dimensional signal. In practical mmWave systems, received signals must be combined in the analog domain, and possibly in the digital domain, before any detection or decoding is performed. For this reason, we revisit the problem of designing practical mmWave receivers in Section \ref{sec:SparseCombining}.

Proceeding with the design of $\Frf\Fbb$, the precoder optimization problem can be stated as
\begin{align}
\begin{split}
(\Frf^\mathrm{opt}, \Fbb^\mathrm{opt})= & \argmax_{\Frf,\ \Fbb}\quad \log_2\left(\left| \bI_{\Ns} + \frac{\rho}{\Ns\sigman^2} \bH \Frf\Fbb \Fbb^*\Frf^*\bH^* \right|\right), \\
& \mathrm{s.t.}\quad \Frf\in \mathcal{F}_\mathrm{RF}, \\ & \hspace{27pt} \|\Frf\Fbb\|^2_{F}=\Ns,
\label{eqn:MutualInformationMaximization}
\end{split}
\end{align}
where $\mathcal{F}_\mathrm{RF}$ is the set of feasible RF precoders, i.e., the set of $\Nt\times \NtRF$ matrices with constant-magnitude entries. To the extent of the authors' knowledge, no general solutions to (\ref{eqn:MutualInformationMaximization}) are known in the presence of the non-convex feasibility constraint $\Frf \in \mathcal{F}_\mathrm{RF}$. Therefore, we propose to solve an approximation of (\ref{eqn:MutualInformationMaximization}) in order to find practical near-optimal precoders that can be implemented in the system of Fig. \ref{fig:MmWaveSystemModel}. 

We start by examining the mutual information achieved by the hybrid precoders $\Frf\Fbb$ and rewriting (\ref{eqn:MutualInformation}) in terms of the ``distance'' between $\Frf\Fbb$ and the channel's optimal unconstrained precoder $\Fopt$. To do so, define the channel's ordered singular value decomposition (SVD) to be $\bH=\bU\SigmaH\bV^*$ where $\bU$ is an $\Nr \times \mathrm{rank}(\bH)$ unitary matrix, $\SigmaH$ is a $\mathrm{rank}(\bH) \times \mathrm{rank}(\bH)$ diagonal matrix of singular values arranged in decreasing order, and $\bV$ is a $\Nt \times \mathrm{rank}(\bH)$ unitary matrix. Using the SVD of $\bH$ and standard mathematical manipulation, (\ref{eqn:MutualInformation}) can be rewritten as
\begin{equation}
\mathcal{I}(\Frf, \Fbb)= \log_2\left(\left| \bI_{\mathrm{rank}(\bH)} + \frac{\rho}{\Ns\sigman^2} \SigmaH^2\bV^* \Frf\Fbb \Fbb^*\Frf^*\bV \right|\right).
\label{eqn:MutualInformationUnitary}
\end{equation}
Further, defining the following two partitions of the matrices $\SigmaH$ and $\bV$ as 
\begin{align}
\begin{split}
\SigmaH= \left[ \begin{array}{cc} \SigmaH_1 & \mathbf{0} \\ \mathbf{0} & \SigmaH_2 \end{array}\right],\qquad \bV=\left[\bV_1\quad \bV_2\right],
\label{eqn:SigmaVPartitions}
\end{split}
\end{align} 
where $\SigmaH_1$ is of dimension $\Ns \times \Ns$ and $\bV_1$ is of dimension $\Nt \times \Ns$, we note that the optimal unconstrained unitary precoder for $\bH$ is simply given by $\Fopt=\bV_1$. Further note that the precoder $\bV_1$ cannot in general be expressed as $\Frf\Fbb$ with $\Frf \in \mathcal{F}_\mathrm{RF}$, and thus cannot be realized in the mmWave architecture of interest. If the hybrid precoder $\Frf\Fbb$ can be made sufficiently ``close'' to the optimal precoder $\bV_1$, however, the mutual information resulting from $\Fopt$ and $\Frf\Fbb$ can be made comparable. In fact, to simplify the forthcoming treatment of $\mathcal{I}(\Frf, \Fbb)$, we make the following system assumption.
\begin{approximation}
We assume that the mmWave system parameters $(\Nt,\Nr,\NtRF,\NrRF)$, as well as the parameters of the mmWave propagation channel $(\Ncl,\Nray,\hdots )$, are such that the hybrid precoders $\Frf\Fbb$ can be made sufficiently ``close'' to the optimal unitary precoder $\Fopt=\bV_1$. Mathematically, this ``closeness'' is defined by the following two equivalent approximations:
\begin{enumerate}
\item The eigenvalues of the matrix $\bI_\Ns - \bV_1^* \Frf\Fbb\Fbb^*\Fbb^*\bV_1 $ are small. In the case of mmWave precoding, this can be equivalently stated as $\bV_1^*\Frf\Fbb \approx \bI_\Ns$.\footnote{For the eigenvalues of $\bI_\Ns-\bV_1^* \Frf\Fbb\Fbb^*\Fbb^*\bV_1$ to be small, we need $\bV_1^*\Frf\Fbb \approx \boldsymbol{\Psi}$ where $\boldsymbol{\Psi}$ is \emph{any $\Ns\times\Ns $ unitary matrix} (not necessarily $\bI_\Ns$). However, if $\Frf\Fbb$ is a valid precoder with $\bV_1^*\Frf\Fbb \approx \boldsymbol{\Psi}$, then so is the rotated precoder $\Frf\widetilde{\bF}_\mathrm{BB}=\Frf\Fbb\boldsymbol{\Psi}^*$ for which we have $\bV_1^*\Frf\widetilde{\bF}_\mathrm{BB}\approx \bI_\Ns $. Since $\Fbb$ can be arbitrarily rotated, the conditions $\bV_1^* \Frf\Fbb\Fbb^*\Fbb^*\bV_1\approx \bI_\Ns $ and $\bV_1^* \Frf \Fbb \approx \bI_\Ns $ can be considered equivalent in this case without loss of generality.}
           
\item The singular values of the matrix $\bV_2^*\Frf\Fbb$ are small; alternatively 
$\bV_2^*\Frf\Fbb \approx \mathbf{0}$.
\end{enumerate}
\label{as:HighResolution}
\end{approximation}
This approximation is similar to the high-resolution approximation used to simplify the analysis of limited feedback MIMO systems by assuming that codebooks are large enough such that they contain codewords that are sufficiently close to the optimal unquantized precoder~\cite{roh2006design}. In the case of mmWave precoding, this approximation is expected to be tight in systems of interest which include: (i) a reasonably large number of antennas $\Nt$, (ii) a number of transmit chains $\Ns < \NtRF\leq \Nt$, and (iii) correlated channel matrices $\bH$. 

Functionally, Approximation \ref{as:HighResolution} allows us further simplify $\mathcal{I}(\Frf, \Fbb)$. To do so, we use the partitions defined in (\ref{eqn:SigmaVPartitions}) and further define the following partition of $\bV^* \Frf\Fbb \Fbb^*\Frf^*\bV$ as 
\begin{equation}
\bV^* \Frf\Fbb \Fbb^*\Frf^*\bV \!=\!\! \left[ \begin{array}{cc} \bV_1^*\Frf\Fbb \Fbb^*\Frf^*\bV_1 & \bV_1^*\Frf\Fbb \Fbb^*\Frf^*\bV_2 \\ \bV_2^*\Frf\Fbb \Fbb^*\Frf^*\bV_1 & \bV_2^*\Frf\Fbb \Fbb^*\Frf^*\bV_2 \end{array} \right]\!\!\! =\!\!\! \left[ \begin{array}{cc} \bQ_{11} & \bQ_{12} \\ \bQ_{21} & \bQ_{22} \end{array}\right], \nonumber
\end{equation}
which allows us to approximate the mutual information achieved by $\Frf\Fbb$ as
\begin{align}
\mathcal{I}(\Frf, \Fbb) & = \log_2\left(\left| \bI_{\mathrm{rank}(\bH)} + \frac{\rho}{\Ns\sigman^2} \SigmaH^2\bV^* \Frf\Fbb \Fbb^*\Frf^*\bV \right|\right) \nonumber \\
& = \log_2\left(\left| \bI_{\mathrm{rank}(\bH)} + \frac{\rho}{\Ns\sigman^2} \left[ \begin{array}{cc} \SigmaH^2_1,& \mathbf{0} \nonumber \\ \mathbf{0} & \SigmaH^2_2 \end{array}\right] \left[ \begin{array}{cc} \bQ_{11} & \bQ_{12} \\ \bQ_{21} & \bQ_{22} \end{array}\right] \right|\right) \nonumber \\
& \stackrel{(a)}{=} \log_2 \left(\left|\bI_\Ns+ \frac{\rho}{\Ns\sigman^2}\SigmaH_1^2\bQ_{11} \right|\right) \nonumber \\ 
& \hspace{15pt} +\log_2 \left(\left|\bI+ \frac{\rho}{\Ns\sigman^2}\SigmaH_2^2\bQ_{22}- \frac{\rho^2}{\Ns^2\sigman^4}\SigmaH_2^2\bQ_{21}\left(\bI_\Ns+\frac{\rho}{\Ns\sigman^2}\SigmaH_1^2\bQ_{11}\right)^{-1}\SigmaH_1^2\bQ_{12}\right|\right) \nonumber \\ 
& \stackrel{(b)}{\approx} \log_2 \left(\left|\bI_\Ns+ \frac{\rho}{\Ns\sigman^2}\SigmaH_1^2\bV_1^*\Frf\Fbb \Fbb^*\Frf^*\bV_1 \right|\right), \label{eqn:MutualInformationSimplified}
\end{align}
where $(a)$ is a result of using the Schur complement identity for matrix determinants and $(b)$ follows from invoking Approximation \ref{as:HighResolution} which implies that $\bQ_{12}$, $\bQ_{21}$ and $\bQ_{22}$ are approximately zero. Using (\ref{eqn:MutualInformationSimplified}), mutual information can be further simplified by writing
\begin{align}
\mathcal{I}(\Frf, \Fbb) & \stackrel{(a)}{\approx} \log_2\left(\left|\bI_\Ns+\frac{\rho}{\Ns\sigman^2}\SigmaH_1^2\right|\right) \nonumber\\ & \hspace{15pt} +\log_2\left(\left|\bI_\Ns-\left(\bI_\Ns+ \frac{\rho}{\Ns\sigman^2}\SigmaH_1^2\right)^{-1} \!\!\! \frac{\rho}{\Ns\sigman^2}\SigmaH_1^2\left(\bI_\Ns \!-\! \bV_1^*\Frf\Fbb \Fbb^*\Frf^*\bV_1\right) \right|\right) \nonumber \\
& \stackrel{(b)}{\approx} \log_2\left(\left|\bI_\Ns+\frac{\rho}{\Ns\sigman^2}\SigmaH_1^2\right|\right)  \nonumber \\ & \hspace{15pt} - \mathrm{tr}\left(\left(\bI_\Ns+ \frac{\rho}{\Ns\sigman^2}\SigmaH_1^2\right)^{-1}\frac{\rho}{\Ns\sigman^2}\SigmaH_1^2\left(\bI_\Ns- \bV_1^*\Frf\Fbb \Fbb^*\Frf^*\bV_1\right)\right) \nonumber \\
& \stackrel{(c)}{\approx} \log_2\left(\left|\bI_\Ns+\frac{\rho}{\Ns\sigman^2}\SigmaH_1^2\right|\right)- \mathrm{tr}\left(\bI_\Ns- \bV_1^*\Frf\Fbb \Fbb^*\Frf^*\bV_1\right) \label{eqn:MutualInformationFinal-1} \\
& = \log_2\left(\left|\bI_\Ns+\frac{\rho}{\Ns\sigman^2}\SigmaH_1^2\right|\right)- \left(\Ns-  \|\bV_1^*\Frf\Fbb\|_{F}^2\right),
\label{eqn:MutualInformationFinal}
\end{align} 
where we note that $(a)$ is exact given (\ref{eqn:MutualInformationSimplified}), and $(b)$ follows from Approximation \ref{as:HighResolution} which implies that the eigenvalues of the matrix $\bX= (\bI_\Ns+ \frac{\rho}{\Ns\sigman^2}\SigmaH_1^2)^{-1}\frac{\rho}{\Ns\sigman^2}\SigmaH_1^2\left(\bI_\Ns- \bV_1^*\Frf\Fbb \Fbb^*\Frf^*\bV_1\right) $ are small and thus allows us to use the following approximation $\log_2 |\bI_\Ns - \bX| \approx \log_2(1-\mathrm{tr}(\bX))\approx -\mathrm{tr}(\bX)$. Finally $(c)$ follows from adopting a high \emph{effective-SNR} approximation which implies that $(\bI+\frac{\rho}{\Ns\sigman^2}\SigmaH_1^2)^{-1}\frac{\rho}{\Ns\sigman^2}\SigmaH_1^2 \approx \bI_\Ns$ and yields the final result in (\ref{eqn:MutualInformationFinal}).\footnote{Note here that it is not the nominal SNR $\frac{\rho}{\Ns\sigman^2}$ that is assumed to be high. This would be a problematic assumption in mmWave systems. It is, however, only the \emph{effective-SNRs} in the channel's dominant $\Ns$ subspaces that are assumed to be sufficiently high. This is a reasonable assumption since these effective SNRs include the large array gain from mmWave beamforming.} We notice that the first term in (\ref{eqn:MutualInformationFinal})  is the mutual information achieved by the optimal precoder $\Fopt=\bV_1$ and that the dependence of $\mathcal{I}(\Frf, \Fbb)$ on the hybrid precoder $\Frf\Fbb$ is now captured in the second and final term of (\ref{eqn:MutualInformationFinal-1}) and (\ref{eqn:MutualInformationFinal}).


Assuming $\Frf\Fbb$ is made exactly unitary, we note that the second term in  (\ref{eqn:MutualInformationFinal-1}) and (\ref{eqn:MutualInformationFinal}) is nothing but the squared chordal distance between the two points $\Fopt=\bV_1$ and $\Frf\Fbb$ on the Grassmann manifold. Since Approximation \ref{as:HighResolution} states the these two points are ``close'', we can exploit the manifold's locally Euclidean property to replace the chordal distance by the Euclidean distance  $\|\Fopt-\Frf\Fbb\|_{F}$ \cite{{lee2012introduction}}. Therefore, near-optimal hybrid precoders that approximately maximize $\mathcal{I}(\Frf, \Fbb)$ can be found by instead minimizing $\|\Fopt-\Frf\Fbb\|_{F}$. In fact, even without treating $\Frf\Fbb$ as a point on the Grassmann manifold, Approximation \ref{as:HighResolution} implies that $\|\bV_1^*\Frf\Fbb\|_{F}^2$, and consequently (\ref{eqn:MutualInformationFinal}), can be approximately maximized by instead maximizing $\mathrm{tr}\left(\bV_1^*\Frf\Fbb\right)$.\footnote{This is since the magnitude of $\bV_1^*\Frf\Fbb$'s off-diagonal entries is negligible and all $\bV_1^*\Frf\Fbb$'s diagonals must be made close to one. Thus $\|\bV_1^*\Frf\Fbb\|_{F}^2$, i.e., the $\ell 2$ norm of $\bV_1^*\Frf\Fbb$'s diagonals, can be maximized by optimizing $\mathrm{tr}\left(\bV_1^*\Frf\Fbb\right)$, i.e., the $\ell 1$ norm of the diagonals~\cite{figiel1977dimension, kwak2008principal, boyd2004convex}.} Since maximizing $\mathrm{tr}\left(\bV_1^*\Frf\Fbb\right)$ is again equivalent to minimizing $\|\Fopt-\Frf\Fbb\|_{F}$, the precoder design problem can be rewritten as
\begin{align}
\begin{split}
(\Frf^\mathrm{opt}, \Fbb^\mathrm{opt})= & \argmin_{\Fbb,\Frf} \|\Fopt-\Frf\Fbb\|_{F},\\
& \mathrm{s.t.}\quad  \Frf\in \mathcal{F}_\mathrm{RF},\\ & \hspace{27pt} \|\Frf\Fbb\|^2_{F}=\Ns,
\label{eqn:ProjectionPrecoding}
\end{split}
\end{align}
which can now be summarized as finding the projection of $\Fopt$ onto the set of hybrid precoders of the form $\Frf\Fbb$ with $\Frf\in \mathcal{F}_\mathrm{RF}$. Further, this projection is defined with respect to the standard Frobenius norm $\|\cdot\|_{F}^2$. Unfortunately, the complex non-convex nature of the feasible set $\mathcal{F}_\mathrm{RF}$ makes finding such a projection both analytically (in closed form) and algorithmically intractable~\cite{tropp2005designing, lewis2008alternating, escalante2011alternating, bauschke1996projection}.

To provide near-optimal solutions to the problem in (\ref{eqn:ProjectionPrecoding}), we propose to exploit the structure of the mmWave MIMO channels generated by the clustered channel model in Section \ref{sec:ChannelModel}. Namely, we leverage the following observations on mmWave precoding:
\begin{enumerate}
\item \emph{Structure of optimal precoder}: Recall that the optimal unitary precoder is $\Fopt=\bV_1$, and that the columns of the unitary matrix $\bV$ form an orthonormal basis for the channel's row space. 
\item \emph{Structure of clustered mmWave channels}: Examining the channel model in (\ref{eqn:ClusteredChannel}), we note that the array response vectors $\at(\Phit_{i\ell},\ \Thetat_{i\ell}), \forall i,\ell,\ \Thetat_{i\ell})$ also form a finite spanning set for the channel's row space. In fact, when $\Ncl\Nray\leq \Nt$, we note that the array response vectors $\at(\Phit_{i\ell},\ \Thetat_{i\ell})$ will be linearly independent with probability one and will thus form another \emph{minimal basis} for the channel's row space when $\Ncl\Nray\leq \min(\Nt,\Nr)$.

\emph{Note:} To establish the linear independence of the vectors $\at(\Phit_{i\ell},\ \Thetat_{i\ell})$, consider the case of uniform linear arrays. When ULAs are considered, the $\Nt \times \Ncl\Nray$ matrix formed by the collection of vectors $\at(\Phit_{i\ell})\ \forall i,\ell$ will be a Vandermonde matrix which has full rank whenever the angles $\Phit_{i\ell}$ are distinct. This event occurs with probability one when $\Phit_{i\ell}$ are generated from a continuous distribution. Linear independence can be established in the case of UPAs by writing their response vectors as a Kronecker product of two ULA response vectors~\cite{el2012capacity}.
\item \emph{Connection between $\Fopt$ and $\at(\Phit_{i\ell}\ \Thetat_{i\ell})$}: Regardless of whether $\Ncl\Nray \leq \Nt$ or not, observation 1 implies that the columns of the optimal precoder $\Fopt=\bV_1$ are related to the vectors $\at(\Phit_{i\ell},\ \Thetat_{i\ell})$ through a linear transformation. As a result, the columns of $\Fopt$ can be written as linear combinations of $\at(\Phit_{i\ell},\ \Thetat_{i\ell}), \forall i,\ell$.
\item \emph{Vectors $\at(\Phit_{i\ell}\ \Thetat_{i\ell})$ as columns of $\Frf$}: Recall that the vectors $\at(\Phit_{i\ell},\ \Thetat_{i\ell})$ are constant-magnitude phase-only vectors which can be applied at RF using analog phase shifters. Therefore, the mmWave transmitter can apply $\NtRF$ of the vectors $\at(\Phit_{i\ell},\ \Thetat_{i\ell})$ at RF (via the RF precoder $\Frf$), and form arbitrary linear combinations of them using its digital precoder $\Fbb$. Namely, it can construct the linear combination that minimizes $\|\Fopt-\Frf\Fbb\|_{F}$.
\end{enumerate}

Therefore, by exploiting the structure of $\bH$, we notice that near-optimal hybrid precoders can be found by further restricting $\mathcal{F}_\mathrm{RF}$ to be the set of vectors of the form $\at(\Phit_{i\ell},\Thetat_{i\ell})$ and solving
\begin{align}
\begin{split}
(\Frf^\mathrm{opt}, \Fbb^\mathrm{opt}) = & \argmin  \|\Fopt-\Frf\Fbb\|_{F},\\
& \mathrm{s.t.} \quad \Frf^{(i)} \in \left\{\at(\Phit_{i\ell},\Thetat_{i\ell}) | \ 1\leq i \leq \Ncl,\ 1\leq \ell \leq \Nray\right\}, \\ 
& \hspace{27pt}   \|\Frf\Fbb\|_{F}^2=\Ns,
\label{eqn:CombinationPrecoding} 
\end{split}
\end{align}
which amounts to finding the best low dimensional representation of $\Fopt$ using the basis vectors $\at(\Phit_{i\ell},\Thetat_{i,\ell})$. We note here that the set of basis vectors can be extended to include array response vectors $\at(\cdot,\cdot)$ in directions other than $\{(\Phit_{i\ell},\Thetat_{i\ell})| \ 1\leq i \leq \Ncl,\ 1\leq \ell \leq \Nray\}$, though the effect of this basis extension is typically negligible.
In any case, the precoding problem consists of selecting the ``best'' $\NtRF$ array response vectors and finding their optimal baseband combination. Finally, we note that the constraint of $\Frf^{(i)}$ can be embedded directly into the optimization objective to obtain the following equivalent problem
\begin{align}
\begin{split}
\widetilde{\bF}_{BB}^\mathrm{opt}=& \argmin_{\widetilde{\bF}_\mathrm{BB}} \|\Fopt-\bA_\mathrm{t}\widetilde{\bF}_\mathrm{BB} \|_{F}, \\
& \mathrm{s.t.}\quad \|\mathrm{diag}(\widetilde{\bF}_\mathrm{BB}\widetilde{\bF}_\mathrm{BB}^*)\|_0=\NtRF, \\ & \hspace{28pt}  \|\bA_\mathrm{t}\widetilde{\bF}_\mathrm{BB}\|_{F}^2=\Ns,
\label{eqn:SparsePrecoding}
\end{split}
\end{align} 
where $\bA_\mathrm{t}=\left[\at(\Phit_{1,1},\Thetat_{1,1}),\ \hdots,\ \at(\Phit_{\Ncl,\Nray},\Thetat_{\Ncl, \Nray})\right]$ is an $\Nt\times \Ncl\Nray$ matrix of array response vectors and $\widetilde{\bF}_\mathrm{BB}$ is an $\Ncl\Nray \times \Ns$ matrix. The matrices $\bA_\mathrm{t}$ and $\widetilde{\bF}_\mathrm{BB}$ act as auxiliary variables from which we obtain $\bF^\mathrm{opt}_\mathrm{RF}$ and $\bF^\mathrm{opt}_\mathrm{BB}$ respectively. Namely, the sparsity constraint $\|\mathrm{diag}(\widetilde{\bF}_\mathrm{BB}\widetilde{\bF}_\mathrm{BB}^*)\|_0=\NtRF$ states that $\widetilde{\bF}_\mathrm{BB}$ cannot have more than $\NtRF$ non-zero rows. When only $\NtRF$ rows of $\widetilde{\bF}_\mathrm{BB}$ are non zero, only $\NtRF$ columns of the matrix $\bA_\mathrm{t}$ are effectively ``selected''. As a result, the baseband precoder $\bF^\mathrm{opt}_\mathrm{BB}$ will be given by the $\NtRF$ non-zero rows of $\widetilde{\bF}_\mathrm{BB}^\mathrm{opt}$ and the RF precoder $\bF^\mathrm{opt}_\mathrm{RF}$ will be given by the corresponding $\NtRF$ columns of $\bA_\mathrm{t}$.

Essentially, we have reformulated the problem of jointly designing $\Frf$ and $\Fbb$ into a sparsity constrained matrix reconstruction problem with one variable. Although the underlying motivation differs, and so does the interpretation of the different variables involved in (\ref{eqn:SparsePrecoding}), the resulting problem formulation is identical to the optimization problem encountered in the literature on sparse signal recovery. Thus, the extensive literature on sparse reconstruction can now be used for hybrid precoder design~\cite{rebollo2002optimized, tropp-signal-recovery}. To see this more clearly, note that in the simplest case of single stream beamforming, (\ref{eqn:SparsePrecoding}) simplifies to
\begin{align}
\begin{split}
\widetilde{\bff}^\mathrm{opt}_{BB} = & \argmin_{\widetilde{\bff}_\mathrm{BB}} \|\bff_\mathrm{opt}-\bA_t\widetilde{\bff}_\mathrm{BB} \|_{F}, \\
& \mathrm{s.t.} \quad \|\widetilde{\bff}_\mathrm{BB}\|_0=\NtRF, 
\quad \|\bA_t\widetilde{\bff}_\mathrm{BB}\|_{F}^2=\Ns,
\label{eqn:SparseBeamforming}
\end{split}
\end{align}
in which the sparsity constraint is now on the vector $\widetilde{\bff}_\mathrm{BB}$. This beamforming problem can be solved, for example, by relaxing the sparsity constraint and using convex optimization to solve its $\ell 2-\ell 1$ relaxation. Alternatively, (\ref{eqn:SparseBeamforming}) can be solved using tools from \cite{tropp2004greed, wright2009sparse, rebollo2002optimized, tibshirani1996regression, tropp-signal-recovery}.


\begin{algorithm}[t!]
\caption{Spatially Sparse Precoding via Orthogonal Matching Pursuit}
\begin{algorithmic}[1]
\REQUIRE $\Fopt$
\STATE $\Frf= \mathrm{Empty\ Matrix}$
\STATE $\bF_\mathrm{res}=\Fopt$
\FOR{$i\leq \NtRF$}
\STATE $\boldsymbol\Psi=\bA_t^*\bF_\mathrm{res}$
\STATE $k= \argmax_{\ell=1,\ \hdots,\ \Ncl\Nray} \left(\boldsymbol\Psi \boldsymbol\Psi^*\right)_{\ell,\ell}$
\STATE $\Frf=\left[\Frf | \bA_t^{(k)} \right]$ 
\STATE $\Fbb=\left(\Frf^*\Frf\right)^{-1}\Frf^*\Fopt$
\STATE $\bF_\mathrm{res}=\frac{\Fopt-\Frf\Fbb}{\|\Fopt-\Frf\Fbb\|_{F}}$
\ENDFOR
\STATE $\Fbb=\sqrt{\Ns} \frac{\Fbb}{\|\Frf\Fbb\|_{F}}$
\RETURN $\Frf,\ \Fbb$
\end{algorithmic}
\label{algo:sparse_SVD}
\end{algorithm}

In the more general case of $\Ns>1$, the problem in (\ref{eqn:SparsePrecoding}) is equivalent to the problem of sparse signal recovery with multiple measurement vectors, also known as the simultaneously sparse approximation problem~\cite{cotter2005sparse, tropp2006algorithmspart1, tropp2006algorithmspart2, chen2006theoretical}. So, for the general case of $\Ns \geq 1$, we present an algorithmic solution based on the well-known concept of orthogonal matching pursuit~\cite{rebollo2002optimized, tropp-signal-recovery, wright2009sparse}. The pseudo-code for the precoder solution is given in Algorithm \ref{algo:sparse_SVD}. In summary, the precoding algorithm starts by finding the vector $\at(\Phit_{i\ell},\Thetat_{i\ell})$ along which the optimal precoder has the maximum projection. It then appends the selected column vector $\at(\Phit_{i\ell},\Thetat_{i\ell})$ to the RF precoder $\Frf$. After the dominant vector is found, and the least squares solution to $\Fbb$ is calculated in step 7, the contribution of the selected vector is removed in step 8 and the algorithm proceeds to find the column along which the ``residual precoding matrix'' $\bF_\mathrm{res}$ has the largest projection. The process continues until all $\NtRF$ beamforming vectors have been selected. At the end of the $\NtRF$ iterations, the algorithm would have (i) constructed an $\Nt \times \NtRF$ RF precoding matrix $\Frf$, and (ii) found the optimal $\NtRF \times \Ns$ baseband precoder $\Fbb$ which minimizes $\|\Fopt-\Frf\Fbb\|_{F}^2$. Step 10 ensures that the transmit power constraint is exactly satisfied.

\begin{figure}[t!]
  \centering
  \subfigure[Beam Pattern of Optimal Beamforming Vector]{
    \includegraphics[width=0.46\linewidth]{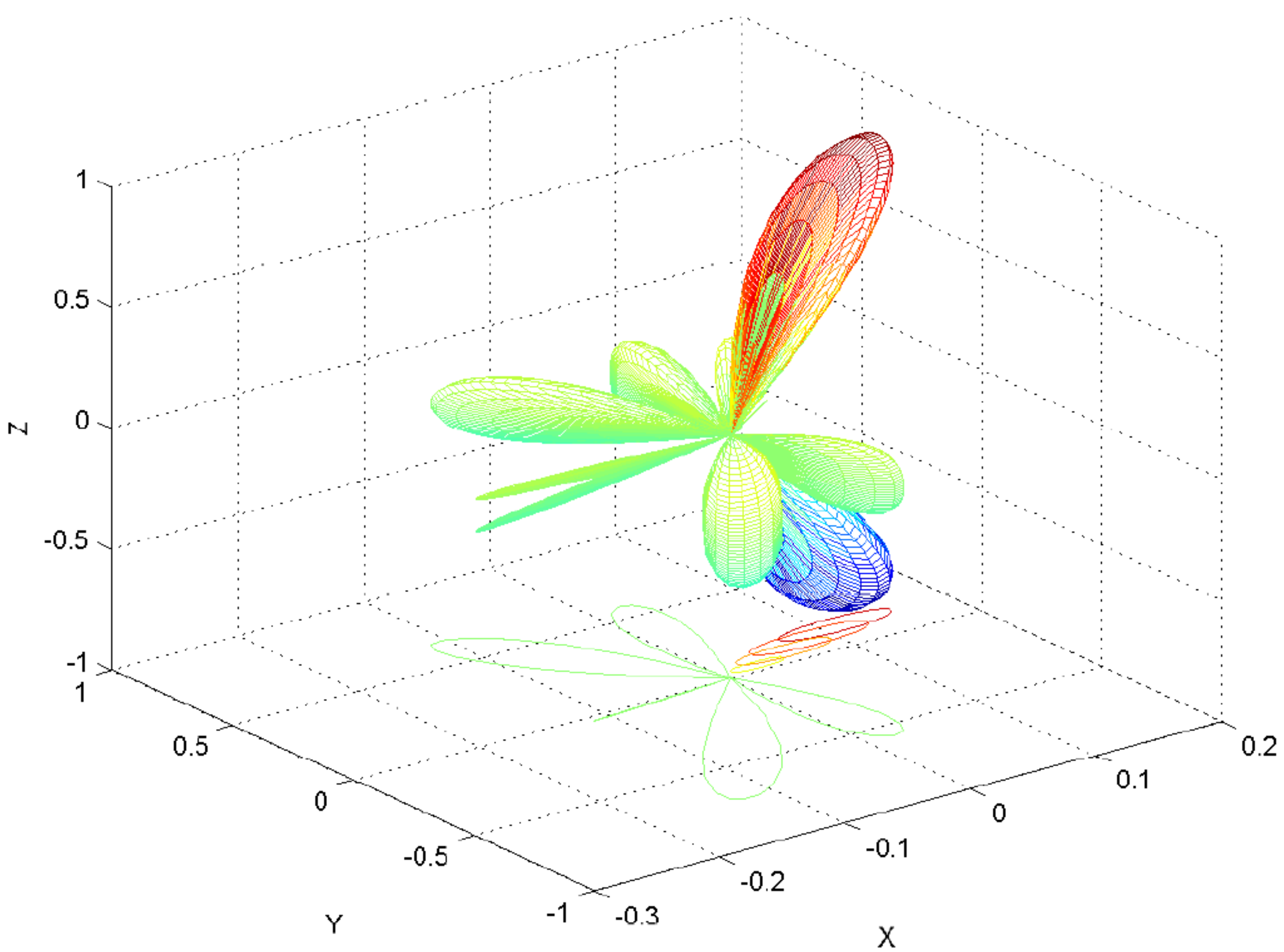}
    \label{subfig:SVDBeamPattern}
  } \hfill
  \subfigure[Beam Pattern with Proposed Solution]{
    \includegraphics[width=0.46\linewidth]{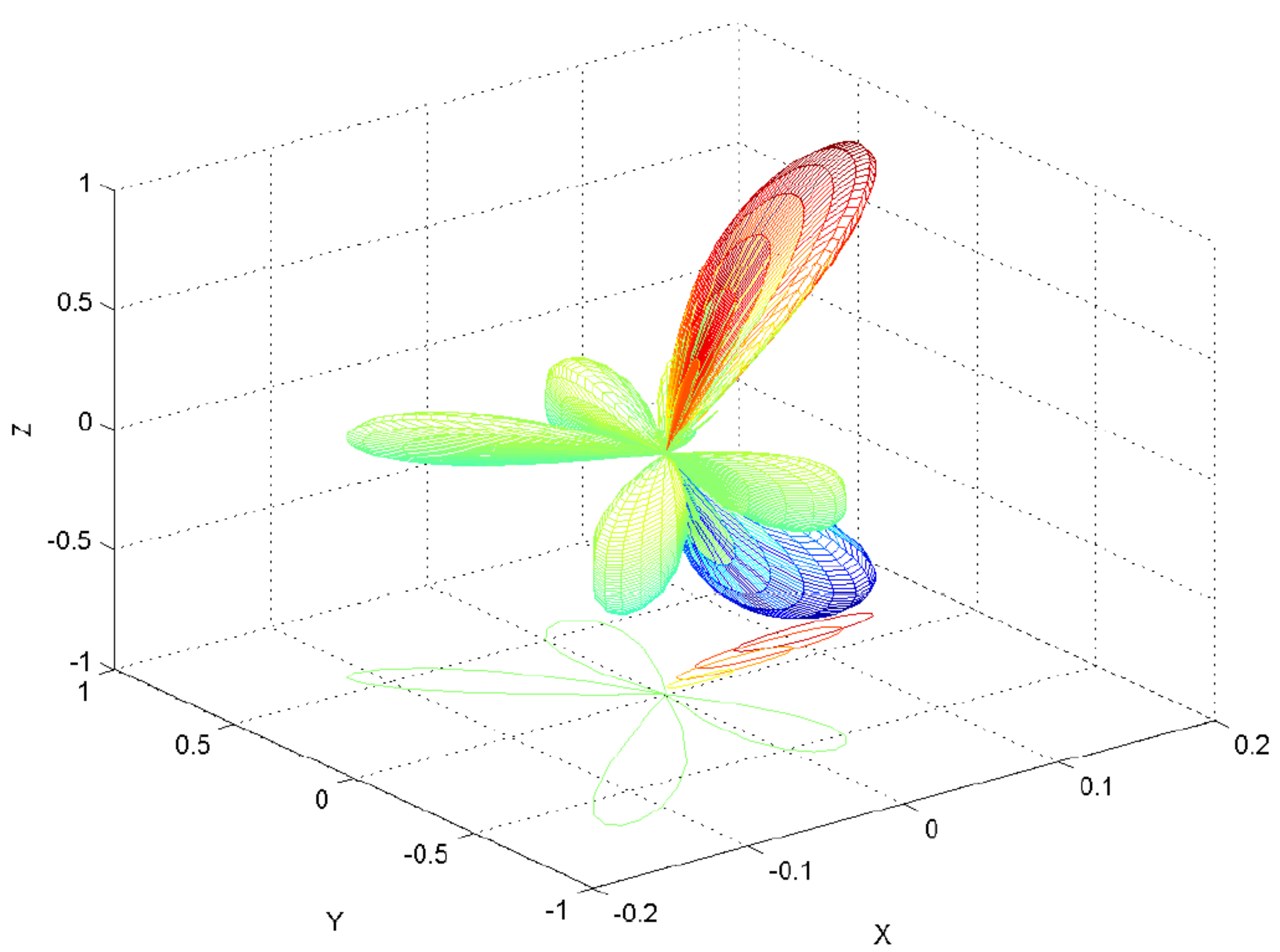}
    \label{subfig:ProposedBeamPattern}
  } \\
  \subfigure[Beam Pattern of Optimal Steering Vector]{
    \includegraphics[width=0.46\linewidth]{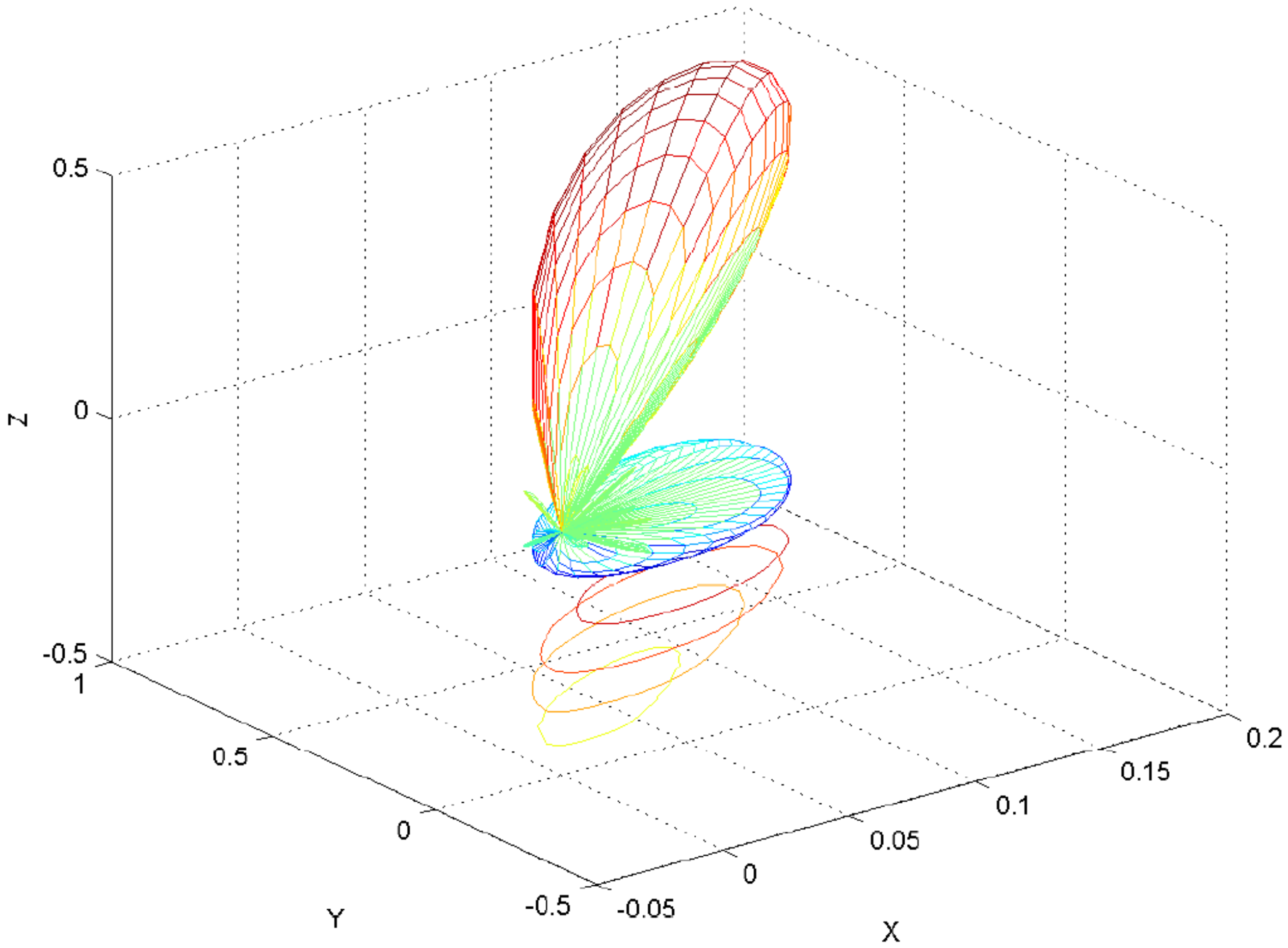}
    \label{subfig:SteeringBeamPattern}
  }
  \caption{Beam pattern generated a 256-element square array in an example channel realization with 6 scattering clusters using (a) optimal unconstrained beamforming, (b) the proposed sparse precoding solution with 4 RF chains, and (c) the beam steering vector in the channel's dominant physical direction. The proposed algorithm is shown to result in beam patterns that closely resemble the patterns generated by optimal beamforming; this beam pattern similarity will ultimately result in similar spectral efficiency. For illustration purposes, the channel's angle spread is set to $0^\circ$ in this figure.}
  \label{fig:BeamPatterns}
\end{figure}

To gain more intuition about the proposed precoding framework, Fig. \ref{fig:BeamPatterns} plots the beam patterns generated by a transmitter with a $256$-element planar array for an example channel realization using (i) the channel's optimal unconstrained precoder, (ii) the proposed precoding strategy with $\NtRF=4$, and (ii) the beam steering vector in the channel's dominant physical direction. We observe that in practical mmWave channels, optimal precoders do in fact generate spatially sparse beam patterns and thus may be accurately approximated by a finite combination of array response vectors. Further, Fig. \ref{fig:BeamPatterns} indicates that Algorithm \ref{algo:sparse_SVD} succeeds in generating beam patterns which closely resemble those generated by $\Fopt$. Therefore, Algorithm \ref{algo:sparse_SVD} succeeds in selecting the best $\NtRF$ steering directions and forming appropriate linear combinations of the selected response vectors. This beam pattern similarity will ultimately result in favorable spectral efficiency performance as shown in Section \ref{sec:Simulations}.

Having presented the proposed precoding framework, we conclude this section with the following design remarks.
\begin{remark}
We note that the mmWave terminals need not know the exact angles $(\Phit_{i\ell},\Thetat_{i\ell})$ that make up the channel matrix $\bH$, and need not use the matrix $\bA_\mathrm{t}$ as defined earlier. We have only used this finite basis for simplicity of exposition. In general, the mmWave terminals can instead select basis vectors of the form $\at(\phi,\theta)$ using any finite set of representative azimuth and elevation directions (such as a set of equally spaced angles for example). This approach avoids having to decompose $\bH$ into its geometric representation and is naturally suited for limited feedback operation. This approach will be discussed further in Section \ref{sec:LimitedFeedback}.
\label{rem:QuantizedAngles}
\end{remark}
\begin{remark}
It may be advantageous in some cases to impose the additional constraint that $\Fbb$ be unitary. Unitary precoders can be more efficiently quantized and are thus more attractive in limited feedback systems. With this additional constraint, (\ref{eqn:SparsePrecoding}) can be solved again via Algorithm \ref{algo:sparse_SVD} by replacing the least squares solution for $\Fbb$ in step 7, by the solution to the corresponding orthogonal Procrustes problem~\cite{gower2004procrustes}. This is given by $\Fbb= \hat{\bU}\hat{\bV}^*$ where $\hat{\bU}$ and $\hat{\bV}$ are unitary matrices defined by the singular value decomposition of $\Frf^*\Fopt$, i.e., $\Frf^*\Fopt=\hat{\bU}\hat{\boldsymbol{\Sigma}}\hat{\bV}^*$~\cite{gower2004procrustes}.
\label{rem:UnitaryFbb}
\end{remark}
\begin{remark}
In the limit of large antenna arrays ($\Nt, \Nr \to \infty$) in very poor scattering environments for which $\Ncl\Nray = o(\min(\Nt,\Nr))$, the results of \cite{el2012capacity, sayeed2010continuous} indicate that simple RF-only beam steering becomes optimal, i.e., it becomes optimal to simply transmit each stream along one of the $\Ns$ most dominant vectors $\at(\Phit_{i\ell},\ \Thetat_{i\ell})$. For arrays of practical sizes, however, Section \ref{sec:Simulations} shows that there can be significant gains from more involved precoding strategies such as the one presented in this section.
\label{rem:AsymptoticBehaviour}
\end{remark}

\section{Practical Millimeter Wave Receiver Design} \label{sec:SparseCombining}

In Section \ref{sec:SparsePrecoding}, we abstracted receiver-side processing and focused on designing practical mmWave precoders that maximize mutual information. Effectively, we assumed that the mmWave receiver can optimally decode data using it $\Nr$-dimensional received signal. Such a decoder can be of prohibitively high complexity in multi-antenna systems, making lower-complexity receivers such as the commonly used linear MMSE receiver more appealing for practical implementation. In fact, in mmWave architectures such as the one shown in Fig. \ref{fig:MmWaveSystemModel}, such optimal decoders are \emph{impossible} to realize since received signals \emph{must} be linearly combined in the analog domain before any detection or decoding is performed. 

In this section, we address the problem of designing linear combiners for the mmWave receiver in Fig. \ref{fig:MmWaveSystemModel}, which uses both analog and digital processing before detection. Assuming the hybrid precoders $\Frf\Fbb$ are fixed, we seek to design hybrid combiners $\Wrf\Wbb$ that minimize the mean-squared-error (MSE) between the transmitted and processed received signals. The combiner design problem can therefore be stated as
\begin{align}
\begin{split}
(\Wrf^\mathrm{opt}, \Wbb^\mathrm{opt})= & \argmin_{\Wrf,\ \Wbb}\quad \bbE\left[\left|\left|\bs-\Wbb^*\Wrf^*\by\right|\right|_2^2\right], \\
& \mathrm{s.t.}\quad \Wrf\in \mathcal{W}_\mathrm{RF}, 
\label{eqn:MMSECombiningProblem}
\end{split}
\end{align}
where $\mathcal{W}_\mathrm{RF}$ is the set of feasible RF combiners, i.e., $\mathcal{W}_\mathrm{RF}$ is the set of $\Nr\times \NrRF$ matrices with constant-gain phase-only entries. In the absence of any hardware limitations that restrict the set of feasible linear receivers, the exact solution to (\ref{eqn:MMSECombiningProblem}) is well known~\cite{kailath2000linear} to be
\begin{align}
\begin{split}
\Wmmse^* & = \bbE\left[\bs\by^*\right]\bbE\left[\by\by^*\right]^{-1} = \frac{\sqrt{\rho}}{\Ns}\Fbb^*\Frf^*\bH^*\left(\frac{\rho}{\Ns} \bH\Frf\Fbb\Fbb^*\Frf^*\bH^* +\sigman^2 \bI_\Nr\right)^{-1} \\
& \stackrel{(a)}{=} \frac{1}{\sqrt{\rho}} \left(\Fbb^*\Frf^*\bH^*\bH\Frf\Fbb +\frac{\sigman^2 \Ns}{\rho} \bI_\Ns\right)^{-1}\Fbb^*\Frf^*\bH^*,
\end{split}
\end{align}
where $(a)$ follows from applying the matrix inversion lemma. Just as in the precoding case, however, this optimal unconstrained MMSE combiner $\Wmmse^*$ need not be decomposable into a product of RF and baseband combiners $\Wbb^*\Wrf^*$ with $\Wrf \in \mathcal{W}_\mathrm{RF}$. Therefore $\Wmmse^*$ cannot be realized in the system of Fig. \ref{fig:MmWaveSystemModel}. Further, just as in the precoding case, the complex non-convex constraint $\Wrf\in \mathcal{W}_\mathrm{RF}$ makes solving (\ref{eqn:MMSECombiningProblem}) analytically impossible and algorithmically non-trivial. To overcome this difficulty, we leverage the methodology used in \cite{michaeli2007constrained, michaeli2008constrained} to find linear MMSE estimators with complex structural constraints.

We start by reformulating the problem in (\ref{eqn:MMSECombiningProblem}) by expanding MSE as follows
\begin{align}
\begin{split}
\bbE\left[\|\bs-\Wbb^*\Wrf^*\by\|_2^2\right] & = \bbE\left[(\bs-\Wbb^*\Wrf^*\by)^*(\bs-\Wbb^*\Wrf^*\by)\right] \\
&= \bbE\left[\mathrm{tr}\left(\left(\bs-\Wbb^*\Wrf^*\by\right)(\bs-\Wbb^*\Wrf^*\by)^*\right)\right] \\
&= \mathrm{tr}\left(\bbE\left[\bs\bs^*\right]\right)-2\Re\left\{\mathrm{tr}\left(\bbE\left[\bs\by^*\right]^*\Wbb^*\Wrf^*\right)\right\} \\ & \hspace{15pt} +\mathrm{tr}\left(\Wbb^*\Wrf^*\bbE\left[\bs\bs^*\right]\Wbb^*\Wrf^*\right).
\end{split}
\end{align}
We now note that since the optimization problem in (\ref{eqn:MMSECombiningProblem}) is over the variables $\Wrf$ and $\Wbb$, we can add any term that is independent of $\Wrf$ and $\Wbb$ to its objective function without changing the outcome of the optimization. Thus, we choose to add the constant term $\mathrm{tr}\left(\Wmmse\bbE\left[\by\by^*\right]\Wmmse^*\right)-\mathrm{tr}\left(\bbE\left[\bs\bs^*\right]\right)$ and minimize the equivalent objective function
\begin{align}
\mathcal{J}(\Wrf,\Wbb)& = \mathrm{tr}\left(\Wmmse\bbE\left[\by\by^*\right]\Wmmse^* \right) -2\Re\left\{\mathrm{tr}\left(\bbE\left[\bs\by^*\right]\Wrf\Wbb \right)\right\} \nonumber \\ & \hspace{15pt} +\mathrm{tr}\left(\Wbb^*\Wrf^*\bbE\left[\bs\bs^*\right] \Wbb^*\Wrf^*\right) \nonumber \\
& \stackrel{(a)}{=} \mathrm{tr}\left(\Wmmse\bbE\left[\by\by^*\right]\Wmmse^*\right) -2\Re\left\{\mathrm{tr}\left(\Wmmse^* \bbE\left[\by\by^*\right]\Wrf\Wbb\right)\right\}  \nonumber \\ & \hspace{15pt} + \mathrm{tr}\left(\Wbb^*\Wrf^*\bbE\left[\bs\bs^*\right]\Wbb^*\Wrf^*\right) \nonumber \\
& = \mathrm{tr}\left(\left(\Wmmse^*-\Wbb^*\Wrf^*\right)\bbE\left[\by\by^*\right]\left(\Wmmse^*-\Wbb^*\Wrf^*\right)^*\right) \nonumber  \\
& = \|\bbE\left[\by\by^*\right]^{1/2}\left(\Wmmse-\Wrf\Wbb\right)\|_{F}^2,
\label{eqn:CombinerObjective}
\end{align}
where $(a)$ follows from noticing that $\Wmmse^*=\bbE\left[\bs\by^*\right]\bbE\left[\by\by^*\right]^{-1}$ which implies that the second term can be rewritten as $\mathrm{tr}\left(\bbE\left[\bs\by^*\right]\Wrf\Wbb\right)=\mathrm{tr}\left(\bbE\left[\bs\by^*\right]\bbE\left[\by\by^*\right]^{-1}\bbE\left[\by\by^*\right]\Wrf\Wbb \right)=\mathrm{tr}\left(\Wmmse^* \bbE\left[\by\by^*\right]\Wrf\Wbb\right)$. As a result of (\ref{eqn:CombinerObjective}), the MMSE estimation problem is equivalent to finding hybrid combiners that solve
\begin{align}
\begin{split}
(\Wrf^\mathrm{opt}, \Wbb^\mathrm{opt})= & \argmin_{\Wrf,\ \Wbb}\quad \|\bbE\left[\by\by^*\right]^{1/2}\left(\Wmmse-\Wrf\Wbb\right)\|_{F} \\
& \mathrm{s.t.}\quad \Wrf\in \mathcal{W}_\mathrm{RF}, 
\label{eqn:MMSEProjectionFormulation}
\end{split}
\end{align}
which amounts to finding the projection of the unconstrained MMSE combiner $\Wmmse$ onto the set of hybrid combiners of the form $\Wrf\Wbb$ with $\Wrf\in \mathcal{W}_\mathrm{RF}$. Thus, the design of MMSE receivers for the mmWave system of interest closely resembles the design of its hybrid precoders. Unlike in the precoding case however, the projection now is not with respect to the standard norm $\|\cdot\|_{F}^2$ and is instead an $\bbE\left[\by\by^*\right]$-weighted Frobenius norm. Unfortunately, as in the case of the precoding problem in (\ref{eqn:ProjectionPrecoding}), the non-convex constraint on $\Wrf$ precludes us from practically solving the projection problem in (\ref{eqn:MMSEProjectionFormulation}). The same observations that allowed us to leverage the structure of mmWave channels to solve the precoding problem in Section \ref{sec:SparsePrecoding}, however, can be translated to the receiver side to solve the combiner problem as well. Namely, because of the structure of clustered mmWave channels, near-optimal receivers can be found by further constraining $\Wrf$ to have columns of the form $\ar(\phi,\theta)$ and instead solving
\begin{align}
\begin{split}
\widetilde{\bW}_\mathrm{BB}^\mathrm{opt}= & \argmin_{\widetilde{\bW}_\mathrm{BB}}\quad \|\bbE\left[\by\by^*\right]^{1/2}\Wmmse-\bbE\left[\by\by^*\right]^{1/2}\bA_\mathrm{r}\widetilde{\bW}_\mathrm{BB}\|_{F}, \\
& \mathrm{s.t.}\quad \|\mathrm{diag}(\widetilde{\bW}_\mathrm{BB}\widetilde{\bW}_\mathrm{BB}^*)\|_0=\NrRF 
\label{eqn:SparseMMSECombining}
\end{split}
\end{align}
where $\bA_\mathrm{r}=\left[\ar(\Phir_{1,1},\Thetar_{1,1}),\ \hdots,\ \at(\Phir_{\Ncl,\Nray},\Thetar_{\Ncl, \Nray})\right]$ is an $\Nr\times \Ncl\Nray$ matrix of array response vectors and $\widetilde{\bW}_\mathrm{BB}$ is an $\Ncl\Nray \times \Ns$ matrix; the quantities $\bA_\mathrm{r}$ and $\widetilde{\bW}_\mathrm{BB}$ act as auxiliary variables from which we obtain $\Wrf$ and $\Wbb$ in a manner similar to Section \ref{sec:SparsePrecoding}.\footnote{As noted in Section \ref{sec:SparsePrecoding} the receiver need not know the exact angles $(\Phir_{i\ell},\Thetar_{i\ell})$ and can instead use any set of representative azimuth and elevation angles of arrival to construct the matrix of basis vectors $\bA_\mathrm{r}$.} As a result, the MMSE estimation problem is again equivalent to the problem of sparse signal recovery with multiple measurement vectors and can thus be solved via the orthogonal matching pursuit concept used in Section \ref{sec:SparsePrecoding}. For completeness the pseudo code is given in Algorithm \ref{algo:sparse_combiner}. 

\begin{algorithm}[t!]
\caption{Spatially Sparse MMSE Combining via Orthogonal Matching Pursuit}
\begin{algorithmic}[1]
\REQUIRE $\Wmmse$
\STATE $\Wrf= \mathrm{Empty\ Matrix}$
\STATE $\bW_\mathrm{res}=\Wmmse$
\FOR{$i\leq \NrRF$}
\STATE $\boldsymbol\Psi=\bA_r^*\bbE\left[\by\by^*\right]\bW_\mathrm{res}$
\STATE $k= \arg\max_{\ell=1,\ \hdots,\ \Ncl\Nray} \left(\boldsymbol\Psi \boldsymbol\Psi^*\right)_{\ell,\ell}$
\STATE $\Wrf=\left[\Wrf | \bA_r^{(k)} \right]$ 
\STATE $\Wbb=\left(\Wrf^*\bbE\left[\by\by^*\right]\Wrf\right)^{-1}\Wrf^*\bbE\left[\by\by^*\right]\Wmmse$
\STATE $\bW_\mathrm{res}= \frac{\Wmmse-\Wrf\Wbb}{\|\Wmmse-\Wrf\Wbb\|_{F}}$
\ENDFOR
\RETURN $\Wrf,\ \Wbb$
\end{algorithmic}
\label{algo:sparse_combiner}
\end{algorithm}

\begin{remark}
This section relaxes the perfect-receiver assumption of Section \ref{sec:SparsePrecoding} and proposes practical methods to find low-complexity linear receivers. The design of precoders and combiners, however, remains decoupled as we have assumed that the precoders $\Frf\Fbb$ are fixed while designing $\Wrf\Wbb$ (and that receivers are optimal while designing $\Frf\Fbb$). This decoupled approach simplifies mmWave transceiver design, and will be shown to perform well in Section \ref{sec:Simulations}, however, some simple ``joint decisions'' may be both practical and beneficial. For example, consider the case where a receiver only has a single RF chain and thus is restricted to applying a single response vector $\ar(\phi,\theta)$. In such a situation, designing $\Frf\Fbb$ to radiate power in $\NtRF$ different directions may lead to a loss in actual received power (since the receiver can only form a beam in one direction). As a result, it is beneficial to account for the limitations of the more-constrained terminal when designing either precoders or combiners. To do so, we propose to run Algorithms \ref{algo:sparse_SVD} and \ref{algo:sparse_combiner} in succession according to the following rules
\begin{align}
\begin{split}
\NtRF & < \NrRF \left\{ \begin{array}{l} \text{1. Solve for } \Frf\Fbb \text{ using Algorithm \ref{algo:sparse_SVD}.}  \\ 
 \text{2. Given } \Frf\Fbb \text{, solve for } \Wrf\Wbb \text{ using Algorithm \ref{algo:sparse_combiner}.} \\ \end{array} \right.  \\
\NtRF & > \NrRF \left\{ \begin{array}{l} \text{1. Solve for } \Wrf\Wbb \text{ using Algorithm \ref{algo:sparse_combiner} assuming } \Frf\Fbb=\Fopt.  \\ 
 \text{2. Solve for } \Frf\Fbb \text{ for the effective channel } \Wbb^*\Wrf^*\bH. \\ \end{array} \right.
\label{eqn:precoder_combiner_order}
\end{split}
\end{align}
\end{remark}
In summary, starting with the more constrained side, the hybrid precoder or combiner is found using Algorithm \ref{algo:sparse_SVD} or \ref{algo:sparse_combiner}. Then, given the output, the remaining processing matrix is found by appropriately updating the effective mmWave channel.

Finally, we note that while the numerical results of Section \ref{sec:Simulations} indicate that this decoupled approach to mmWave transceiver design yields near-optimal spectral efficiency, a more direct joint optimization of $(\Frf, \Fbb,\Wrf,\Wbb)$ is an interesting topic for future investigation. Similarly, while we have solved the sparse formulation of the precoding and combining problems via orthogonal matching pursuit, the problems in (\ref{eqn:SparsePrecoding}) and (\ref{eqn:SparseMMSECombining}) can be solved by leveraging other algorithms for simultaneously sparse approximation~\cite{tropp2006algorithmspart2}.

\section{Limited Feedback Spatially Sparse Precoding} \label{sec:LimitedFeedback}

Section \ref{sec:SparsePrecoding} implicitly assumed that the transmitter has perfect knowledge of the channel matrix $\bH$ and is thus able to calculate $\Fopt$ and approximate it as a hybrid RF/baseband precoder $\Frf\Fbb$. Since such transmitter channel knowledge may not be available in practical systems, we propose to fulfill this channel knowledge requirement via limited feedback~\cite{love-heath-EGT, love-heath-grassmannian-beamforming, love-heath-limited-feedback, roh2006design}. Namely, we assume that the receiver (i) acquires perfect knowledge of $\bH$, (ii) calculates $\Fopt$ and a corresponding hybrid approximation $\Frf\Fbb$, then (iii) feeds back information about $\Frf\Fbb$ to the transmitter. Since hybrid precoders are naturally decomposed into an RF and baseband component, we propose to quantize $\Frf$ and $\Fbb$ separately while exploiting the mathematical structure present in each of them.

\subsection{Quantizing the RF Precoder} \label{sec:RFQuantization}

Recall that the precoder $\Frf$ calculated Section \ref{sec:SparsePrecoding} has $\NtRF$ columns of the form $\at(\phi, \theta)$. Therefore, $\Frf$ admits a natural parametrization in terms of the $\NtRF$ azimuth and elevation angles that it uses. Thus, $\Frf$ can be efficiently encoded by quantizing its $2\NtRF$ free variables. 
For simplicity, we propose to uniformly quantize the $\NtRF$ azimuth and elevation angles using $\Nphi$ and $\Ntheta$ bits respectively. Therefore, the quantized azimuth and elevation angles are such that 
\begin{align}
\begin{split}
\hat{\phi}_k & \in \mathcal{C}_\phi =\left\{\Phit_\mathrm{min}+ \frac{\Phit_\mathrm{max}-\Phit_\mathrm{min}}{2^{\Nphi+1}},\ \Phit_\mathrm{min}+ \frac{3(\Phit_\mathrm{max}-\Phit_\mathrm{min})}{2^{\Nphi+1}},\ \hdots,\ \Phit_\mathrm{max}- \frac{\Phit_\mathrm{max}-\Phit_\mathrm{min}}{2^{\Nphi+1}} \right\}
\\
\hat{\theta}_k & \in \mathcal{C}_\theta =\left\{\Thetat_\mathrm{min}+ \frac{\Thetat_\mathrm{max}-\Thetat_\mathrm{min}}{2^{\Ntheta+1}},\ \Thetat_\mathrm{min}+ \frac{3(\Thetat_\mathrm{max}-\Thetat_\mathrm{min})}{2^{\Ntheta+1}},\ \hdots,\ \Thetat_\mathrm{max}- \frac{\Thetat_\mathrm{max}-\Thetat_\mathrm{min}}{2^{\Ntheta+1}} \right\}
\label{eqn:SteeringCodebooks}
\end{split}
\end{align}
where we recall that $[\Phit_\mathrm{min}, \ \Phit_\mathrm{max}]$ and $[\Thetat_\mathrm{min}, \ \Thetat_\mathrm{max}]$ are the sectors over which $\Lambda_\mathrm{t}(\phi,\theta)\neq 0$. The receiver can then quantize $\Frf$ by simply selecting the entries of $\mathcal{C}_\phi$ and $\mathcal{C}_\theta$ that are closest in Euclidean distance to $\Frf$'s angles. Alternatively, as stated in Remark \ref{rem:QuantizedAngles}, Algorithm \ref{algo:sparse_SVD} can be run directly using the $\Nt \times 2^{\Nphi+\Ntheta}$ matrix of ``quantized response vectors''
\begin{equation}
\bA^\mathrm{quant.}_\mathrm{t}=\left[\at(\Phit_1,\Thetat_1),\ \hdots,\ \at(\Phit_i,\Thetat_\ell),\ \hdots,\ \at(\Phit_{2^\Nphi},\Thetat_{2^\Ntheta})\right],
\end{equation}
and the index of the selected angles can be fed back to the transmitter. While this latter approach has higher search complexity, it has the advantage of (i) ``jointly quantizing'' all $2\NtRF$ angles, and (ii) automatically matching the baseband precoder $\Fbb$ to the quantized angles.

\subsection{Quantizing the Baseband Precoder} \label{sec:BBQuantization}

To efficiently quantize $\Fbb$, we begin by highlighting its mathematical structure in mmWave systems of interest. Namely, we note that for systems with large antenna arrays, we typically have that $\Frf^*\Frf\approx \bI_\NtRF$. When coupled with Approximation \ref{as:HighResolution}, we have that $\Fbb^*\Fbb\approx \bI_\Ns$, i.e., $\Fbb$ is approximately unitary. In fact, $\Fbb$ can be made exactly unitary as discussed in Remark \ref{rem:UnitaryFbb}. Further, we recall that the spectral efficiency expression in (\ref{eqn:SpectralEfficiency}) is invariant to $\Ns\times\Ns$ unitary transformations of the baseband precoder. Therefore, $\Fbb$ is a subspace quantity that can be quantized on the Grassmann manifold~\cite{love-heath-grassmannian-beamforming, love-heath-limited-feedback}. Suitable codebooks for $\Fbb$ can be designed using Lloyd's algorithm on a training set of baseband precoders and using the chordal distance as a distance measure~\cite{gray_vec_quant}.
Since such codebook construction is well-studied in the literature on limited feedback MIMO, we omit its details for brevity and refer the reader to \cite[Section IV]{zhou2006ber} for an in-depth description of the process.
%

\section{Simulation Results} \label{sec:Simulations}
\begin{figure} [t]
  \centering
  \includegraphics[width=4.5in]{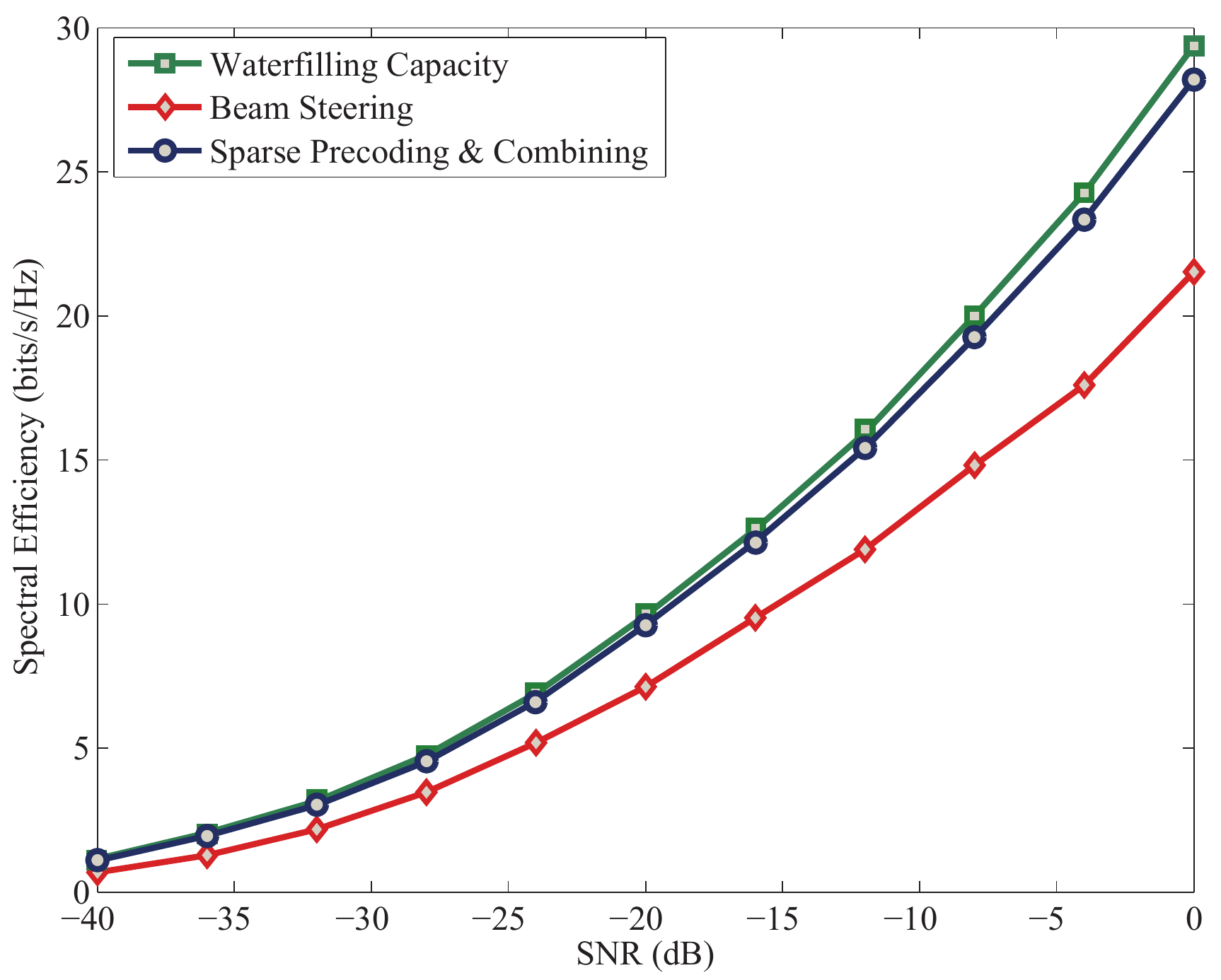}
  \caption{This figure compares the spectral efficiency achieved when rank adaptation and unequal power allocation is allowed in $256\times 64$ system with $\NtRF= \NrRF=4$. It is shown that sparse precoding and combining can approach the performance of an unconstrained capacity-achieving (waterfilling) precoder. The figure also demonstrates large gains over a beam steering strategy in which streams are sent along different physical directions with a similar unequal power allocation.}
  \label{fig:Capacity_4x4rf_256x64UPA_sectored_Low_SNR}
  \vspace{-0.2in}
\end{figure}

In this section, we present simulation results to demonstrate the performance of the spatially sparse precoding algorithm presented in Section \ref{sec:SparsePrecoding} when combined with the sparse MMSE combining solution presented in Section \ref{sec:SparseCombining}. We model the propagation environment as a $\Ncl=8$ cluster environment with $\Nray=10$ rays per cluster with Laplacian distributed azimuth and elevation angles of arrival and departure~\cite{xu2002spatial, forenza2007simplified}. For simplicity of exposition, we assume all clusters are of equal power, i.e., $\sigma^2_{\alpha,i}=\sigma^2_{\alpha}\ \forall i$, and that the angle spread at both the transmitter and receiver are equal in the azimuth and elevation domain, i.e., $\sigma_\Phit = \sigma_\Phir = \sigma_\Thetat = \sigma_\Thetar$. Since outdoor deployments are likely to use sectorized transmitters to decrease interference and increase beamforming gain, we consider arrays of directional antenna elements with a response given in (\ref{eqn:AntennaElementGain})~\cite{pi2011introduction, hendrantoro2002use}. The transmitter's sector angle is assumed to be $60^\circ$-wide in the azimuth domain and $20^\circ$-wide in elevation~\cite{pi2011introduction}. In contrast, we assume that the receivers have relatively smaller antenna arrays of omni-directional elements; this is since receivers must be able to steer beams in any direction since their location and orientation in real systems is random. The inter-element spacing $d$ is assumed to be half-wavelength. We compare the performance of the proposed strategy to optimal unconstrained precoding in which streams are sent along the channel's dominant eigenmodes. We also compare with a simple beam steering solution in which data streams are steered onto the channel's best propagation paths.\footnote{Note that, when $\Ns > 1$, the best propagation paths in terms of spectral efficiency may not be the ones with the highest gains. This is since, with no receiver baseband processing, different paths must be sufficiently separated so as they do not interfere. In this case, the best paths are chosen via a costly exhaustive search. Further, when power allocation is considered in Fig. \ref{fig:Capacity_4x4rf_256x64UPA_sectored_Low_SNR}, the same waterfilling power allocation is applied to the beam steering solution.} For fairness, the same total power constraint is enforced on all precoding solutions and signal-to-noise ratio is defined as $\text{SNR}=\frac{\rho}{\sigman^2}$.

\begin{figure} [t]
  \centering
  \includegraphics[width=4.5in]{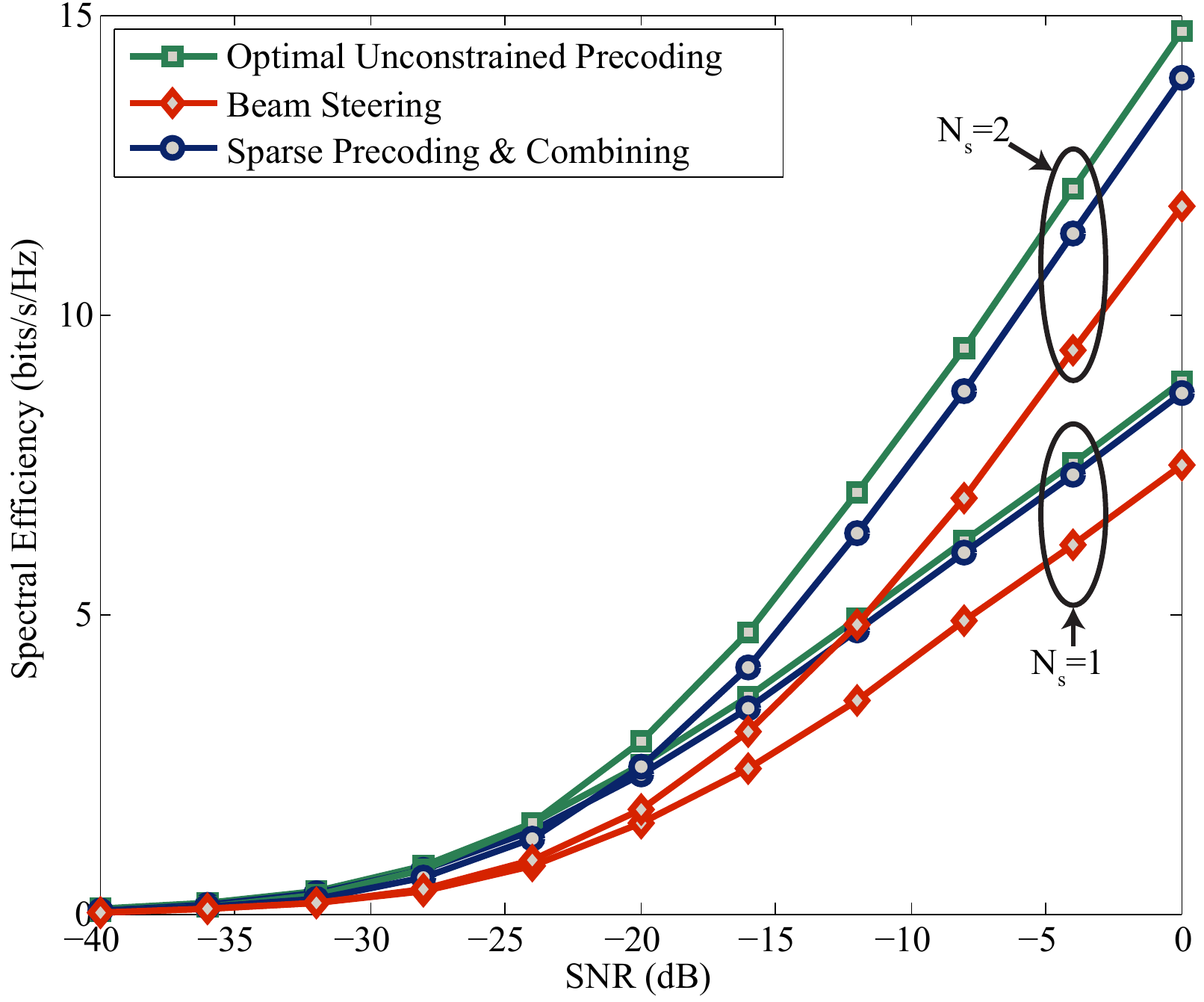}
  \caption{Spectral Efficiency achieved by various precoding solutions for a $64\times 16$ mmWave system with planar arrays at the transmitter and receiver. The propagation medium is a $\Ncl=8$ cluster environment with $\Nray=10$ and an angular spread of $7.5^\circ$. Four RF chains are assumed to be available for sparse precoding and MMSE combining.}
  \label{fig:RateVsSNR_64x16UPA_4RFChain_6Cluster75Spread}
  \vspace{-0.2in}
\end{figure}

\begin{figure} [t]
  \centering
  \includegraphics[width=4.5in]{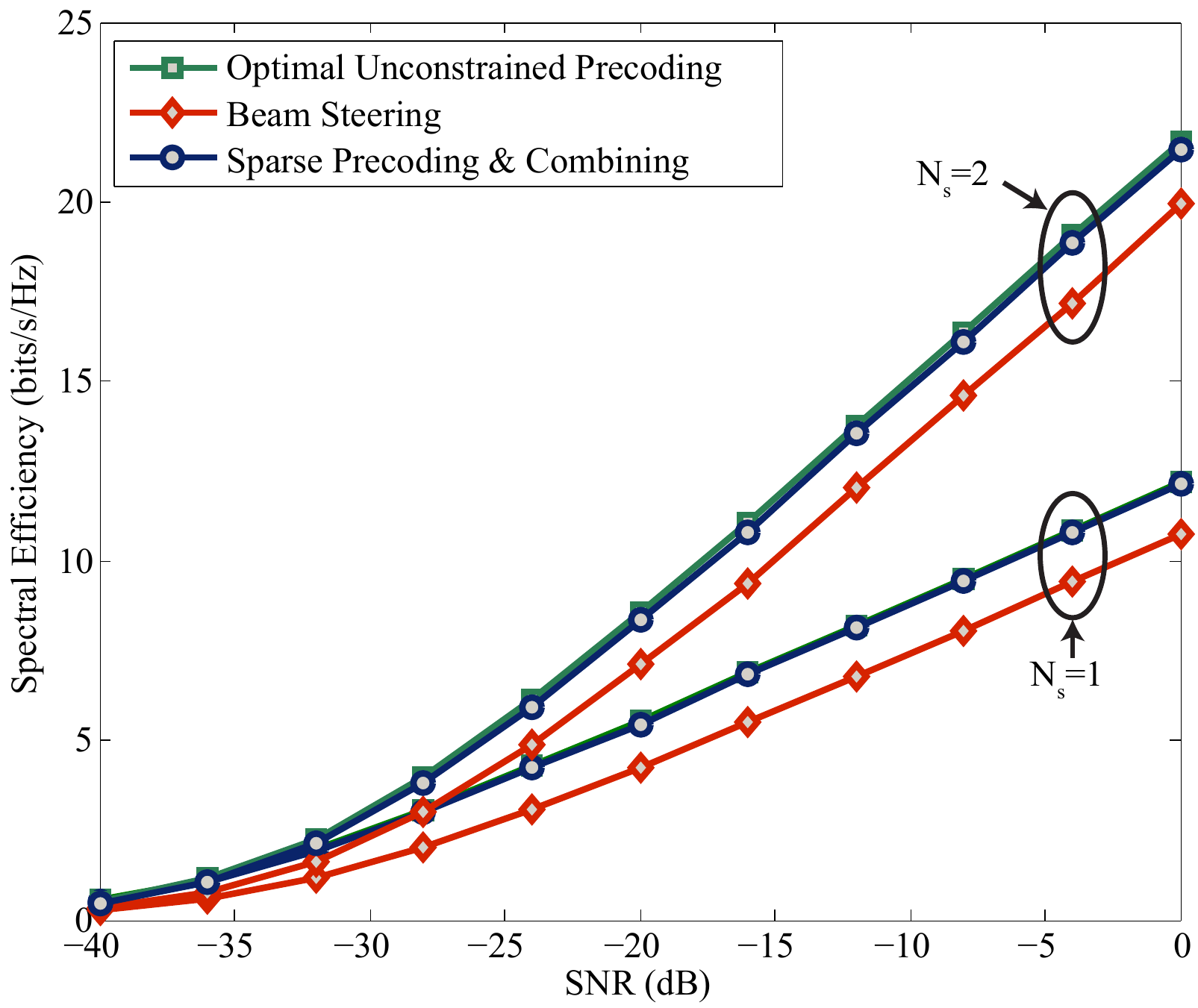}
  \caption{Spectral Efficiency achieved in a $256\times 64$ mmWave system with planar arrays at the transmitter and receiver. Channel parameters are set as in Fig. \ref{fig:RateVsSNR_64x16UPA_4RFChain_6Cluster75Spread}. Six RF chains are available for sparse precoding and combining.}
  \label{fig:RateVsSNR_256x64UPA_6RFChain_6Cluster75Spread}
  \vspace{-0.2in}
\end{figure}

Fig. \ref{fig:RateVsSNR_64x16UPA_4RFChain_6Cluster75Spread} shows the spectral efficiency achieved in a $64\times 16$ system with square planar arrays at both transmitter and receiver. For the proposed precoding strategy, both transmitter and receiver are assumed to have four transceiver chains with which they transmit $\Ns=1$ or $2$ streams. Fig. \ref{fig:RateVsSNR_64x16UPA_4RFChain_6Cluster75Spread} shows that the proposed framework achieves spectral efficiencies that are essentially equal to those achieved by the optimal unconstrained solution in the case $\Ns=1$ and are within a small gap from optimality in the case of $\Ns=2$. This implies that the proposed strategy can very accurately approximate the channel's dominant singular vectors as a combination of four steering vectors. When compared to traditional beam steering, Fig. \ref{fig:RateVsSNR_64x16UPA_4RFChain_6Cluster75Spread} shows that there is a non-negligible improvement to be had from more sophisticated precoding strategies in mmWave systems with practical array sizes. To explore performance in mmWave systems with larger antenna arrays, Fig. \ref{fig:RateVsSNR_256x64UPA_6RFChain_6Cluster75Spread} plots the performance achieved in a $256 \times 64$ system with $\NtRF=\NrRF=6$ RF chains. Fig. \ref{fig:RateVsSNR_256x64UPA_6RFChain_6Cluster75Spread} shows that the proposed precoding/combining solution achieves almost-perfect performance in both $\Ns=1$ and $\Ns=2$ cases. Further, we note that although beam steering is expected to be optimal in the limit of large arrays, as discussed in Remark \ref{rem:AsymptoticBehaviour}, the proposed solution still outperforms beam steering by approximately 5 dB in this larger mmWave system.

While Section \ref{sec:SparsePrecoding} focused on the design of fixed-rank precoders with equal power allocation across streams, the same framework can be applied to systems in which $\Ns$ is determined dynamically and streams are sent with unequal power. This configuration allows us to compare the rates achieved by the proposed precoding/combining framework to the mmWave channel's waterfilling capacity. To do so, Algorithm \ref{algo:sparse_SVD} is simply set to approximate $\Fopt=\bV\boldsymbol{\Gamma}$ where $\boldsymbol{\Gamma}$ is a diagonal matrix resulting from the waterfilling power allocation. Fig. \ref{fig:Capacity_4x4rf_256x64UPA_sectored_Low_SNR} demonstrates the performance achieved when Algorithms \ref{algo:sparse_SVD} and \ref{algo:sparse_combiner} are used to approximate the channel's capacity-achieving precoders and combiners in a $256 \times 64$ mmWave system with $\NtRF=\NrRF=4$. Fig. \ref{fig:Capacity_4x4rf_256x64UPA_sectored_Low_SNR} shows that the proposed framework allows systems to approach channel capacity and provides large gains over simple beam steering. Since the multiplexing gain of the mmWave system is limited by $\Ns\leq \min\{\NtRF,\NrRF\}$, capacity cannot be approached at very high SNR when the optimal $\Ns$ exceeds $\min\{\NtRF,\NrRF\}$. Fig. \ref{fig:Capacity_4x4rf_256x64UPA_sectored_Low_SNR} indicates, however, that even at an SNR of 0 dB where we observe that $\Ns=3$ streams are sent over most channel realizations, the proposed strategy is still within a small gap from capacity. Finally, we note that although the derivation leading up to (\ref{eqn:MutualInformationFinal}) does not account for unequal power allocation across streams, Fig. \ref{fig:Capacity_4x4rf_256x64UPA_sectored_Low_SNR} indicates that Algorithm \ref{algo:sparse_SVD} is nevertheless a sensible approach to designing such precoders.

\begin{figure} [t]
  \centering
  \includegraphics[width=4.5in]{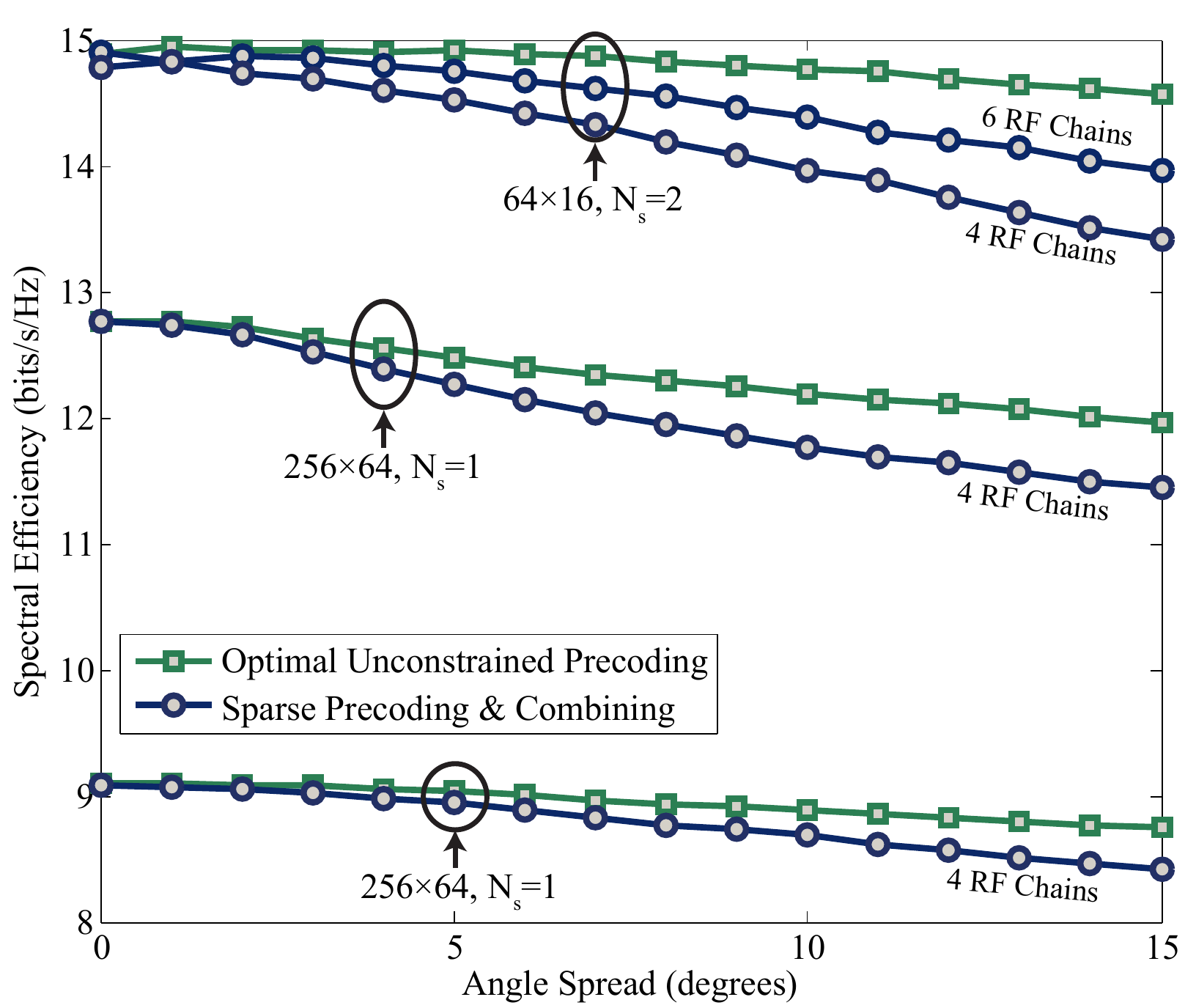}
  \caption{Spectral Efficiency vs. Angle Spread in a number of different mmWave system configuration at an SNR of 0 dB. For simplicity of exposition, we assume that the angle spread is such that $\sigma_\Phit = \sigma_\Phir = \sigma_\Thetat = \sigma_\Thetar$. It is shown that as angle spread increased, and scattering becomes richer, the performance of the proposed algorithm degrades. However, the rate gap remains below $10\%$ at a significant angle spread of $15^\circ$. For more reasonable angle spreads of around $5^\circ$, the rate gap is negligible.}
  \label{fig:RateVsAngleSpread_SNR=0db}
  \vspace{-0.2in}
\end{figure}

The proposed precoding/combining framework leverages the mathematical structure of large mmWave channels with relatively limited scattering. To examine performance in propagation environments with varying levels of scattering, Fig. \ref{fig:RateVsAngleSpread_SNR=0db} plots spectral efficiency as a function of the channel's angle spread for a number of mmWave system configurations. Fig. \ref{fig:RateVsAngleSpread_SNR=0db} indicates that when the angle spread is low, i.e., the scattering is rather limited, the performance of the proposed algorithm is within a small gap from the performance of unconstrained precoding. As angle spread increases, the rates achieved by the proposed solutions slowly degrade. However, Fig \ref{fig:RateVsAngleSpread_SNR=0db} indicates that in the two $\Ns=1$ cases shown, the rate gap remains below $10\%$ at a significant angle spread of $15^\circ$ and is negligible for more reasonable angle spreads of around $5^\circ$. In the case of $\Ns > 1$ with smaller arrays, spectral efficiency degrades more rapidly with angle spread. This can be seen by examining the $64 \times 16$ system with $\NtRF=\NrRF=4$ and $\Ns=2$. If possible, the effect of increased scattering can be mitigated by increasing the number of RF chains at the mmWave terminals which enables them to generate more flexible precoders/combiners. This can be seen by examining the same $64 \times 16$ system with $\NtRF=\NrRF=6$.

\begin{figure} [t]
  \centering
  \includegraphics[width=4.5in]{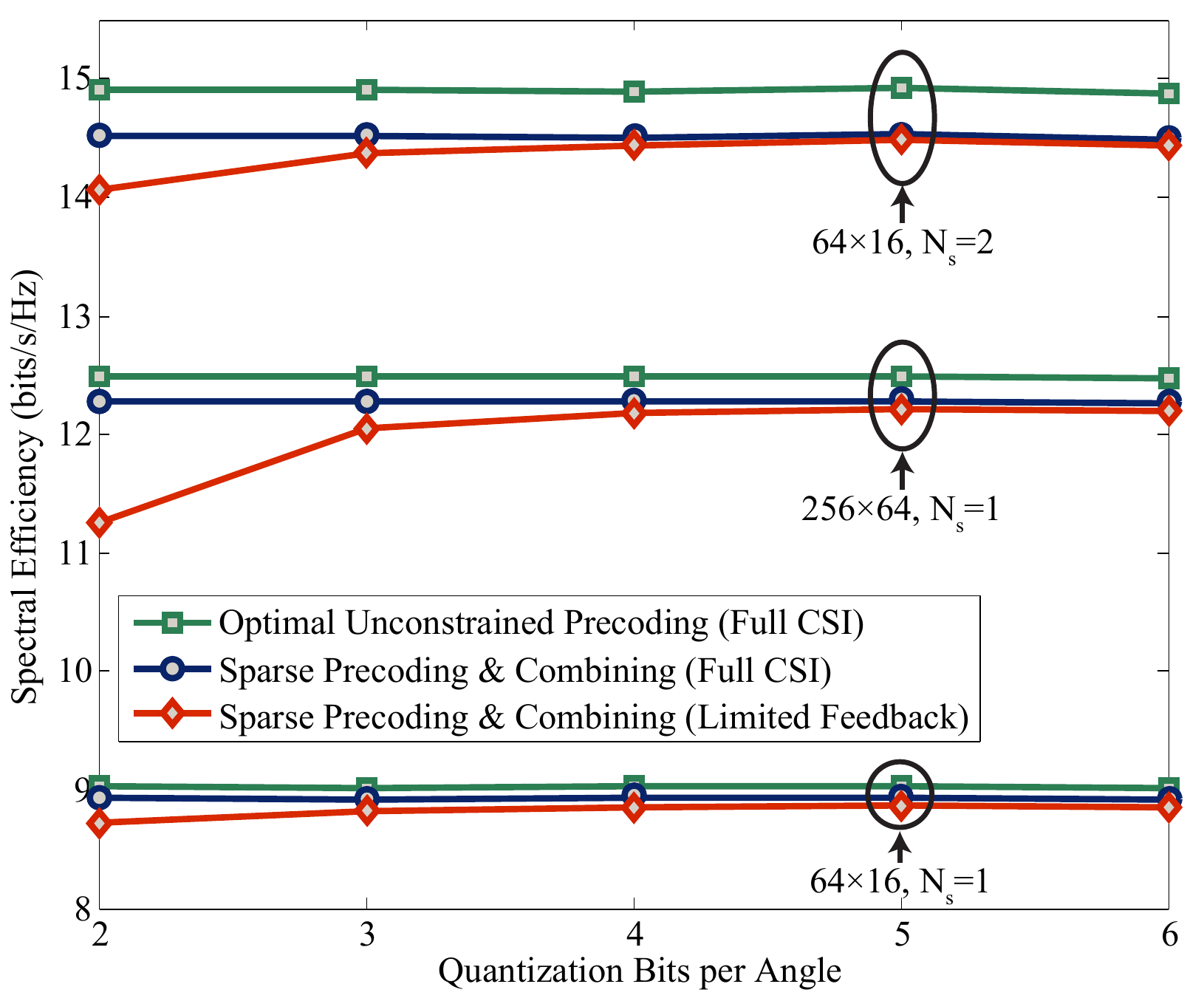}
  \caption{Spectral Efficiency vs. Quantization Bits per Angle different mmWave system configurations, all with $\NtRF= \NrRF=4$, at an SNR of 0 dB. For simplicity of exposition, we assume that $\Nphi=\Ntheta$ and an baseband precoder codebook of 4 bits in the $\Ns=1$ case and 6 bits in the $\Ns=2$ case. The figure indicates that for the considered array sizes, 3 bits per angle is often enough to achieve almost-perfect performance.}
  \label{fig:RateVsQuantizationBits_SNR=0db}
  \vspace{-0.2in}
\end{figure}

Finally, we examine the performance of the proposed precoding strategy in systems without channel state information at the transmitter. For this performance characterization, we assume that the receiver calculates $\Frf$ and $\Fbb$ with full knowledge of the channel and feeds back their parameters as described in Section \ref{sec:LimitedFeedback}. We assume that the receiver uses four and six bits to quantize $\Fbb$ in the case of $\Ns=1$ and $\Ns=2$ respectively, and constructs codebooks as described in Section \ref{sec:BBQuantization}. The receiver uses a variable number of bits to quantize the azimuth and elevation angles used in $\Frf$. For simplicity of exposition, we assume that $\Nphi=\Ntheta$. Fig. \ref{fig:RateVsQuantizationBits_SNR=0db} indicates that similar performance can be expected in limited feedback systems and that the performance degradation due to quantization is limited. Namely, Fig. \ref{fig:RateVsQuantizationBits_SNR=0db} indicates that no more than 3 bits are needed to quantize each steering angle in practical systems, and even 2 bits yields almost-perfect performance for a $64\times 16$ systems with $\Ns=1$. In general the number of bits needed to properly quantize the steering angles grows slowly with array size since larger arrays generate narrower beams and require finer steering. Since beam width is inversely proportional to the antenna array dimensions, a reasonable rule-of-thumb is to add 1 bit per azimuth (or elevation) steering angle whenever the array's width (or height) doubles. Fig. \ref{fig:RateVsQuantizationBits_SNR=0db} is promising as it indicates that it takes no more than 20 bits to quantize a $64 \times 1$ precoder and about 22 bits for a $64 \times 2$ precoder. When considering the fact that practical mmWave systems will use twenty to fifty times more antennas compared to traditional MIMO systems, which use about 4 to 6 bits of feedback~\cite{love-heath-limited-feedback}, we see that exploiting spatial sparsity in precoding helps dramatically compress feedback and keep its overhead manageable.

\section{Conclusion} \label{sec:Conclusion}
In this paper we considered single user precoding and combing in mmWave systems where traditional MIMO solutions are made infeasible by the heavy reliance on RF precoding. By leveraging the structure of realistic mmWave channels, we developed a low hardware-complexity precoding solution. We formulated the problem of mmWave precoder design as a sparsity-constrained signal recovery problem and presented an algorithmic solution using orthogonal matching pursuit. We showed that the same framework can be applied to the problem of designing practical MMSE combiners for mmWave systems. We showed that the proposed precoders can be efficiently quantized and that the precoding strategy is well-suited for limited feedback systems. Finally, we presented numerical results on the performance of spatially sparse mmWave processing and showed that it allows systems to approach their theoretical limits on spectral efficiency. Future work related to such mmWave precoding includes relaxing the assumptions made throughout this paper such as (i) perfect channel state information at the receiver, (ii) knowledge of the antenna array structure, and (iii) the specialization to narrowband channels.

%
%


\singlespacing
\bibliographystyle{IEEEtran}
\bibliography{IEEEabrv,sparse-beamforming}

\newpage

\end{document}

%% file: input.tex
\usepackage{amsfonts}
\usepackage{times}
\usepackage{latexsym}
\usepackage{dsfont}
\usepackage{amssymb}
\usepackage{amsmath}
\usepackage{cite}
\usepackage{verbatim}
\usepackage{subfigure}

\newtheorem{theorem}{Theorem}

\newtheorem{approximation}[theorem]{Approximation}

\newtheorem{remark}[theorem]{Remark}

\newcommand{\Ns}{N_s}

\def\bb0{{\mathbb{0}}}

\def\ba{{\mathbf{a}}}
\def\bb{{\mathbf{b}}}

\def\bff{{\mathbf{f}}}

\def\bn{{\mathbf{n}}}

\def\bs{{\mathbf{s}}}

\def\bx{{\mathbf{x}}}
\def\by{{\mathbf{y}}}

\def\b0{{\mathbf{0}}}

\def\bA{{\mathbf{A}}}
\def\bB{{\mathbf{B}}}

\def\bF{{\mathbf{F}}}

\def\bH{{\mathbf{H}}}
\def\bI{{\mathbf{I}}}

\def\bQ{{\mathbf{Q}}}
\def\bR{{\mathbf{R}}}

\def\bU{{\mathbf{U}}}
\def\bV{{\mathbf{V}}}
\def\bW{{\mathbf{W}}}
\def\bX{{\mathbf{X}}}

\def\bbE{{\mathbb{E}}}

\def\sf0{{\mathsf{0}}}

\def\Nt{{N_t}}
\def\Nr{{N_r}}

\def\Ns{{N_s}}

